\documentclass[11pt]{article}

\usepackage{fullpage}
\usepackage{amssymb}
\usepackage{amsmath}
\usepackage{graphics}
\usepackage{graphicx}
\usepackage{boxedminipage}
\usepackage{xy}
\xyoption{all}

\newtheorem{theorem}{\bf Theorem}[section]

\newtheorem{lemma}[theorem]{\bf Lemma}

\newtheorem{claim}[theorem]{\bf Claim}
\newtheorem{observation}[theorem]{\bf Observation}

\newcommand{\qed} {\hfill$\Box$}

\newcommand{\eat}[1] {}
\newcommand{\proof} {{\it Proof: }}
\newcommand{\calP} {{\cal P}}
\newcommand{\calA} {{\cal A}}
\newcommand{\calE} {{\cal E}}
\newcommand{\calT} {{\cal T}}
\newcommand{\calH} {{\cal H}}
\newcommand{\calD} {{\cal D}}
\newcommand{\Acc} {{\rm Acc}}
\newcommand{\Inst} {{\rm Inst}}
\newcommand{\wh}[1] {\widehat{#1}}
\newcommand{\wt}[1] {\widetilde{#1}}
\newcommand{\Opt} {{\tt Opt}}
\newcommand{\LCA} {{\rm LCA}}
\newcommand{\mypath} {{\rm path}}
\newcommand{\val} {{\rm val}}
\newcommand{\mypred} {{\rm pred}}
\newcommand{\mysucc} {{\rm succ}}
\newcommand{\TMIS} {{\rm Time(MIS)}}
\newcommand{\pair}[2] {\langle #1,#2 \rangle}
\newcommand{\BuildBTD} {{\tt BuildBalTD}}
\newcommand{\BuildITD} {{\tt BuildIdealTD}}
\newcommand{\ceil}[1] {\lceil #1 \rceil}
\newcommand{\floor}[1] {\lfloor #1 \rfloor}
\newcommand{\rt} {{\rm rt}}
\newcommand{\dl} {{\rm dl}}
\newcommand{\len} {{\rm len}}
\newcommand{\depth} {{\rm depth}}
\newcommand{\mymid} {{\rm mid}}
\newcommand{\hmin} {h_{\min}}

\title{Distributed Algorithms for Scheduling on Line and Tree Networks} 
\author{
Venkatesan T. Chakaravarthy \and Sambuddha Roy \and Yogish Sabharwal
}
\date{IBM Research Lab, New Delhi, India\\
\{vechakra,sambuddha,ysabharwal\}@in.ibm.com
}

\begin{document}

\maketitle              

\begin{abstract}
We have a set of processors (or agents) and a set of graph networks defined over some vertex set.
Each processor can access a subset of the graph networks.
Each processor has a demand specified as a pair of vertices $\pair{u}{v}$, along with a profit; 
the processor wishes to send data between $u$ and $v$.
Towards that goal, the processor needs to select a graph network accessible to it
and a path connecting $u$ and $v$ within the selected network.
The processor requires exclusive access to the chosen path, in order to route the data.
Thus, the processors are competing for routes/channels.
A feasible solution selects a subset of demands and schedules each selected demand
on a graph network accessible to the processor owning the demand; the solution also 
specifies the paths to use for this purpose.
The requirement is that for any two demands scheduled on the same graph network,
their chosen paths must be edge disjoint.
The goal is to output a solution having the maximum aggregate profit.
Prior work has addressed the above problem in a distibuted setting
for the special case where all the graph networks are simply paths (i.e, line-networks).
Distributed constant factor approximation algorithms are known for this case.

The main contributions of this paper are twofold. 
First we design a distributed constant factor approximation algorithm for the more general case of tree-networks.
The core component of our algorithm is a tree-decomposition technique, which may be of independent interest.
Secondly, for the case of line-networks, we improve the known approximation guarantees by a factor of $5$.
Our algorithms can also handle the capacitated scenario, wherein the demands and edges have bandwidth requirements
and capacities, respectively.
\end{abstract}

\section{Introduction}
Consider the following fundamental scheduling/routing problem.
We have a set $V$ consisting of $n$ points or vertices.
A set of $r$ undirected graphs provide communication networks over these vertices.
All the edges in the graphs provide a uniform bandwidth, say $1$ unit.
There are $m$ processors (or agents) each having access to a subset of the communication networks.
Each processor $P$ has a demand/job $a$ specified as a pair of vertices $u$ and $v$,
and a bandwidth requirement (or {\em height}) $h(a)\leq 1$.
The processor $P$ wishes to send data between $u$ and $v$,
and for this purpose, the processor can use any of the networks $G$ accessible to it.
To send data over a network $G$, the processor $P$ 
requires a bandwidth of $h(a)$ along some path (or route) connecting the pair of vertices 
$u$ and $v$ in $G$. The input specifies a profit for each demand.
A feasible solution is to select a subset of demands and schedule each selected demand on some graph-network.
For each selected demand $\pair{u}{v}$ scheduled on a graph-network $G$,
the feasible solution must also specify which path connecting $u$ and $v$ must be used for transmission.
The following conditions must be satisfied:
(i) {\em Accessibility requirement: }If a demand $\pair{u}{v}$ owned by a processor $P$
is scheduled on a graph-network $G$, then $P$ should be able to access $G$;
(ii) {\em Bandwidth requirement: }For any network $G$ and for any edge $e$ in $G$, 
the sum of bandwidth requirements of selected demands that use the edge $e$ must not exceed $1$ unit 
(the bandwidth offered by the edge).
We call this the {\em throughput maximization problem}\footnote{The generalization in which the bandwidths offered by edges can vary has also been studied.
For the case where there is only one graph, this is known as the {\em unsplittable flow problem} (UFP),
which has been well-studied (see survey \cite{UFPSurvey}). 
In this paper, we shall only consider the case where the bandwidth offered by all the edges are uniform, say $1$ unit}.
We shall refer to the special case of the problem wherein the heights of all demands is $1$ unit 
as the {\em unit height case}.
In this case, we see that the paths of any two demands scheduled on the same network should be edge disjoint. 
The general case wherein the heights can be arbitrary will be referred
to as the {\em arbitrary height case}.

It is known that the throughput maximization problem is 
NP-hard to approximate within a factor of $\Omega(\log^{1/2-\epsilon} n)$, even for the unit height case of a 
single graph-network \cite{EDP-Hardness}.
Constant factor approximations are known for special cases of the throughput maximization problem (c.f. \cite{AmitKumar}).
Our goal in this paper is to study the problem in a distributed setting.
Prior work has addressed the problem in a distributed setting for the special case of line networks.
In our paper, we present distributed algorithms for the more general case of tree networks
and also improve the known approximation ratios for the case of line networks.
We first discuss the concept of line networks and summarize the known sequential and distributed algorithms
for this case. \\

\begin{figure}
\centering
\includegraphics[width=4.5in]{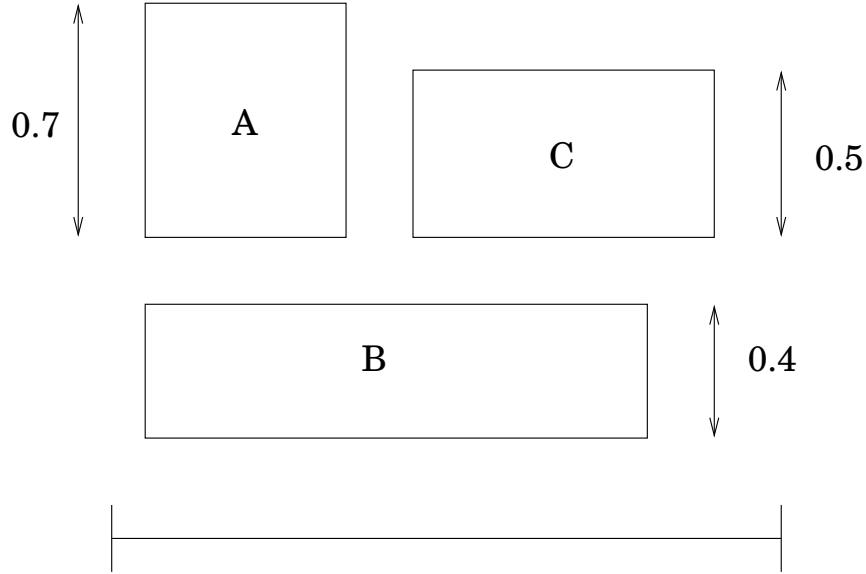}
\caption{Illustration for the problem on line-networks. 
The bandwidth/capacity offered by the resource is $1$ unit throughout the timeline. 
The sets of demands $\{A,C\}$ and $\{B,C\}$ can be scheduled
on the resource, but both $A$ and $B$ cannot be scheduled on the same resource.
}
\label{fig:1}
\end{figure}

\noindent
{\bf Line-Networks:} A {\em line-network} refers to a graph which is simply a path.
Consider the special case of the throughput maximization problem
wherein all the graph-networks are identical paths; say the path is $1,2,\ldots, n$.
We can reformulate this special case by viewing the path as a timeline.
We visualize each edge $(i,i+1)$ as a timeslot so that the number of timeslots
is $n-1$, say numbered $1,2,\ldots,n-1$; then the timeline consisting of these
timeslots becomes a range $[1,n-1]$.
Each demand pair $\pair{u}{v}$ can be represented by the timeslots $u,u+1, \ldots,v-1$
and can be viewed as a interval $[u,v-1]$. 
Thus, each demand can be assumed to be specified as an interval $[s,e]$,
where $s$ and $e$ are the starting and ending timeslots.
Each graph network can be viewed as a resource offering a uniform bandwidth of $1$ unit
throughout the timeline. We see that a feasible solution selects a set of 
demands and schedules each demand on a resource accessible to the processor owning
the demand such that for any resource and any timeslot, 
the sum of heights of the demands scheduled on the resource and active at the timeslot
does not exceed $1$ unit. The goal is to choose a subset of demands with the maximum throughput.
See Figure \ref{fig:1} for an illustration.

In natural applications, a demand may specify a {\em window} 
$[\rt,\dl]$ (release time and deadline) where it can be executed and a processing time $\rho$.
The job can be executed on any time segment of length $\rho$ contained within the window.
The rest of the problem description remains the same as above.
In the new setup, apart from selecting a set of demands and determining the resources where they must be executed,
a feasible solution must also choose a execution segment for each selected demand.
As before, the accessibility and the bandwidth constraints must be satisfied.
The goal is to find a feasible solution having maximum profit.

The throughput maximization problem on line-networks 
has been well-studied in the realm of classical, sequential computation. 
For the arbitrary height case, Bar-Noy et al. \cite{Bar-Noy-Jacm} presented a $5$-approximation algorithm.
For the unit height case, Bar-Noy et al. \cite{Bar-Noy-Jacm}, and independently Berman and Dasgupta \cite{BermanDasgupta}
presented $2$-approximation algorithms; both these algorithms can also handle the notion of windows.
Generalizations and special cases of the problem have also been studied\footnote{
For the case where there is only one line-network and there are no windows,
improved approximations are known \cite{Bar-Noy-Jacm,Calinescu}.
The UFP problem on line-networks (where the bandwidth offered varies over the timeline)
has also been well studied (see \cite{Bansal1,Bansal2,our-ipdps,AmitKumar,ChekuriUFP})
and a constant factor approximation algorithm is known \cite{Bonsma}.
}.

Panconesi and Sozio \cite{Pancc,Pancj} studied the throughput maximization problem on line-networks 
in a distributed setting. In this setup, two processors can communicate with each other, 
if they have access to some common resource.
We shall assume the standard synchronous, message passing model of computation:
in a given network of processors, each processor can communicate in one step with all
other processors it is directly connected to. The running time of the algorithm is given
by the number of communication rounds. This model is universally used in the context
of distributed graph algorithms. We require that the local computation at any processor
takes only polynomial time. To be efficient, we require the communication rounds
to be polylogarithmic in the input size. 
We can construct a communication graph taking the processors to be the vertices
and drawing an edge between two processors, if they can communicate (i.e., they share a common resource).
Notice that the diameter of the communication graph can be as large
as the number of processors $m$. 
So, there may be a pair of processors such that the path connecting them has a large number of hops (or edges).
Hence, within the stipulated polylogarithmic number of rounds, it would be infeasible to
send information between such a pair of processors.
The above fact makes it challenging to design distributed algorithms with polylogarithmic number of rounds.

Under the above model, Panconesi and Sozio \cite{Pancj} designed distributed approximation
algorithms for the throughput maximization problem on line networks.
For the case of unit height demands, they presented an algorithm with an approximation ratio of $(20+\epsilon)$
(throughout the paper, $\epsilon>0$ is a constant fixed arbitrarily).
For the general arbitrary height case, they devised an algorithm with an approximation ratio of $(55+\epsilon)$.
Both the above algorithms can also handle the notion of windows.
The number of communication rounds of these algorithms is:
$O\left(\frac{\TMIS}{\epsilon\cdot h_{\min}} \log{\frac{L_{\max}}{L_{\min}}} \log{\frac{p_{\max}}{p_{\min}}}\right)$.
Here, $L_{\max}$ and $L_{\min}$ are the maximum and minimum length of any demand,
and $p_{\max}$ and $p_{\min}$ are the maximum and minimum profit of any demand.
The value $h_{\min}$ is the minimum height of any demand (recall that all demand heights are at most $1$ unit);
in the case of unit height demands, $h_{\min}=1$.
The value $\TMIS$ is the number of rounds needed for computing a maximal independent set (MIS) in general graphs.
The randomized algorithm of Luby \cite{Luby} can compute MIS in $O(\log N)$ rounds,
where $N=nmr$ ($n$, $m$ and $r$ are the number of timeslots, demands and resources, respectively);
if this algorithm is used, then the overall distributed algorithm would also be randomized.
Alternatively, via network-decompositions, \cite{PancSrini} present a deterministic algorithm 
with $\TMIS = O(2^{\sqrt{\log N}})$.\\

\begin{figure}[t]
\centering
\includegraphics[width=4.5in]{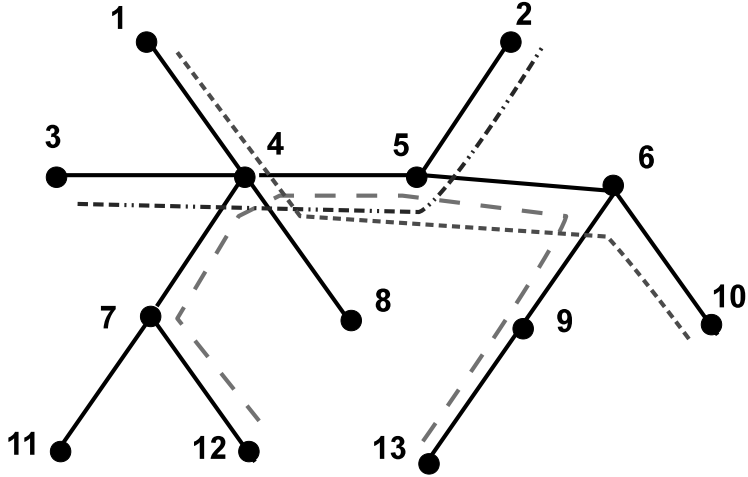}
\caption{
Tree-Networks: There are three demands $\pair{1}{10}$, $\pair{2}{3}$ and $\pair{12}{13}$.
In the unit height case, only one of the three demands can be scheduled on the given tree-network
(because they all share the edge $\pair{4}{5}$).
To illustrate the arbitrary height case, suppose their heights are $0.4$, $0.7$ and $0.3$, respectively.
Then, the first and third demand can be scheduled together.
}
\label{fig:2}
\end{figure}

\noindent
{\bf Our Contributions: }In this paper, we make two important contributions. 
The first is that we provide improved approximation ratios
for the throughput maximization problems on line-networks addressed by Panconesi and Sozio \cite{Pancj}. 
Secondly, we present distributed approximation algorithms for the more general case of tree-networks.
A {\em tree-network} refers to a graph which is a tree. 
Notice that in a tree, the path between a pair of vertices $u$ and $v$ is unique
and so, it suffices if the feasible solution schedules each selected demand on a tree-network
and the paths will be determined uniquely (see Figure \ref{fig:2}).

Prior work has addressed the throughput maximization problem for the scenario where
the input consists of a single tree-network (and all processors have access to the sole tree-network).
Under this setup, Tarjan showed that the unit height case can be solved in polynomial time \cite{Tarjan}.
Lewin-Eytan et al. \cite{Lewin-Eytan} presented a $5$-approximation algorithm for the arbitrary height case.
In the setting of multiple tree-networks, the problem is NP-hard even for the unit height case.
By extending the algorithm of Lewin-Eytan et al., we can show that the problem 
can be approximated within a factor of $3$ and $8$, for the unit height and arbitrary height cases, respectively.

One of the main goals of the current paper is to design distributed algorithms
for the throughput maximization problems on tree-networks. 
Our main result is:
\begin{quote}
{\bf Main result}: We present a distributed $(7+\epsilon)$-approximation algorithm for the unit height case of the 
throughput maximization problem on tree-networks. 
\end{quote}
The number of communication rounds is polylogarithmic in the input size:
$O(\TMIS \cdot (1/\epsilon) \cdot \log n \cdot \log(p_{\max}/p_{\min}))$.
Here, $n$ is the number of vertices;
$p_{\max}$ and $p_{\min}$ are the maximum and minimum profits.
$\TMIS$ is the number of rounds taken for computing MIS in arbitrary graphs with $N$ vertices,
where $N=mr$ ($m$ is the number of processors/demands and $r$ is the number of input tree-networks).
As in the work of Panconesi and Sozio \cite{Pancj}, the size of each message is
$O(M)$ where $M$ is the number of bits needed for encoding the information about a demand 
(such as its profit, end-points and height).

Recall that Panconesi and Sozio \cite{Pancj} presented a 
distributed $(20+\epsilon)$-approximation algorithm for the unit height case of the line-networks problem. 
The main result provides improvements over the above work along two dimensions:
the new algorithm can handle the more general concept of tree-networks and simultaneously, it
offers an improved approximation ratio. 

Extending the main result, we design a distributed $(80+\epsilon)$-approximation algorithm for  the
arbitrary height case of the tree-networks problem
The number of communication rounds taken by this algorithm is
$O(\TMIS \cdot (1/\epsilon) \cdot (1/h_{\min}) \cdot \log n \cdot \log(p_{\max}/p_{\min}))$.
This algorithm assumes that the value $h_{\min}$ is known to all the processors.
Alternatively, we assume that a value $h_{\min}$ is fixed a priori
and all the demands are required to have height at least $h_{\min}$.

Next, we provide a improved approximation ratios
for the case of line-networks with windows. We design distributed algorithms
with approximation ratios $(4+\epsilon)$ and $(23+\epsilon)$, for the unit height case
and arbitrary height case, respectively
\footnote{The conference version of the paper \cite{podc} claimed approximation ratios of $(20+\epsilon)$
and $(11+\epsilon)$ for the arbitrary height case of tree and line networks, respectively.
However, there was a minor error in analyzing the approximation guarantee of the algorithm.
The error is fixed in the current paper with an increase in the ratios}.
The number of communication rounds taken by these algorithms is the same as that
of Panconesi and Sozio \cite{Pancj}. 

\noindent
{\bf Proof Techniques and Discussion: }
At a technical level, our paper makes two main contributions. 
The algorithms of Panconesi and Sozio \cite{Pancj}, as well as our algorithms,
go via the primal-dual method (see \cite{ShmoysBook}). The sequential algorithms of Bar-Noy et al. \cite{Bar-Noy-Jacm}
and Lewin-Eytan et al. \cite{Lewin-Eytan} use the local ratio technique, but they
can also be reformulated as primal-dual algorithms. Given a demand/job,
there are multiple tree-networks (or line-networks) where the demand can be scheduled and we call each such possibility
as a demand instance. All of the above algorithms work in two phases: 
in the first phase, a subset of candidate demand instances are identified
and an assignment to dual variables is computed. In the second phase,
the candidate set is pruned and a feasible solution is constructed.
The dual assignment is used as a lowerbound for the optimal solution,
by appealing to the weak-duality theorem. In fact, approximation algorithms
for many other packing problems utilize the above two-phase strategy.

We first formulate the above two-phase method as a framework.
An important feature of the framework is that any algorithm following the framework
must produce an ordering of the demand instances and also for each demand instance,
it must determine the edges along the path whose dual variables will be increased (or {\em raised}).
The ordering and the chosen edges should satisfy a certain property called the ``interference property".
The number of edges chosen, denoted $\Delta$, is a factor in determining the approximation ratio.
In the case of line-networks, Panconesi and Sozio \cite{Pancj} 
classify the demand instances into logarithmic many groups based on their lengths
and obtain an ordering with $\Delta=3$.
In the case of tree-networks, it is more challenging to design an ordering
satisfying the interference property. Towards that goal, we introduce the notion of
``tree-decompositions". 
The efficacy of a tree-decomposition is measured by its
depth and ``pivot size" $\theta$. As it turns out, the pivot size $\theta$ determines the parameter $\Delta$
and the depth determines the number of rounds taken by the algorithm. 
Our first main technical contribution is a tree-decomposition with depth $O(\log n)$ and
pivot size $\theta=2$. Using this tree-decomposition, we show how to get an ordering with $\Delta=6$.
Our tree-decompositions may be of independent interest.

Another feature of the framework is that an algorithm following the framework should
produce an assignment for the dual variables in the first phase.
This assignment need not form a dual feasible solution, but it should be approximately feasible:
the dual assignment divided by a parameter $\lambda$ ($0< \lambda \leq 1$) should yield a
feasible solution. The approximation ratio is inversely related to the parameter $\lambda$.
The algorithm of Panconesi and Sozio \cite{Pancj} produces a dual assignment with parameter 
$\lambda=1/(5+\epsilon)$. Our second main technical contribution is a method
for constructing a dual assignment with parameter $\lambda=(1-\epsilon)$.
Thus, we get a improved approximation ratios for the case
of line-networks.

\section{Unit Height Case of Tree Networks: Problem Definition}
The input consists of a vertex set $V$ containing $n$ vertices,
a set of $m$ processors $\calP$, a set of $m$ {\em demands} $\calA$ and
a set of $r$ {\em tree-networks} $\calT$ (each defined over the vertex-set $V$).
A demand $a\in \calA$ is specified as a pair of vertices $a=(u,v)$ 
and it is associated with a profit $p(a)$;
$u$ and $v$ are called the {\em end-points} of $a$.
Each processor $P\in \calP$ owns a unique demand $a\in \calA$.
For each processor $P\in \calP$, the input also provides a set $\Acc(P)\subseteq \calT$
that specifies the set of tree-networks {\em accessible} to $P$.
Let $p_{\max}$ and $p_{\min}$ be the maximum and minimum profits.
We will assume that all the tree-networks are connected. 
Note that the tree-networks can have different sets of edges and so, they are allowed to define different trees.

A feasible solution $S$ selects a set of demands $S\subseteq \calA$
and schedules each $a\in S$ on some tree-network $T\in \calT$.
The feasible solution must satisfy the following properties:
(i) for any $a\in S$, if $a$ is owned by a processor $P$
and $a$ is scheduled on a tree-network $T$, then $P$ must be able to access $T$ (i.e., $T\in \Acc(P)$);
(ii) for any two selected demands $a_1=(u_1, v_1)$ and $a_2=(u_2, v_2)$,
if both $a_1$ and $a_2$ are scheduled on the same tree-network $T$,
then the path between $u_1$ and $v_1$, and the path between $u_2$ and $v_2$ in the
tree-network $T$ must be edge-disjoint (meaning, the two paths must not share any edge).
The profit of solution $S$ is defined to be the sum of profits of the selected demands;
this is denoted $p(S)$. The problem is to find the maximum profit feasible solution.

We next present a reformulation of the problem, which will be more convenient for our discussion.
Consider each demand $a\in \calA$ and let $P$ be the processor which owns $a$.
For each tree-network $T\in \Acc(P)$, create a copy of $a$ with the same end-points and profit; 
we call this the {\em demand instance} of $a$ belonging to the tree-network $T$.
Let $\calD$ denote the set of all demand instances over all the demands;
each demand instance $d\in \calD$ can represented by its two end-points and the tree-network to which it belongs.
For a demand $a$ owned by a processor $P$, 
let $\Inst(a)$ denote the set of all instances of $a$ (we have $|\Inst(a)|=|\Acc(P)|$). 
The profit of a demand instance $d\in \calD$ is defined to be the same
as that of the demand to which it belongs; we denote this as $p(d)$.
A feasible solution selects a subset of demand instances $S\subseteq \calD$
such that: 
(i) for any two demand instances $d_1,d_2\in S$, if $d_1$ and $d_2$ belong to the same tree-network $T$,
then their paths (in the tree-network $T$) do not share any edge;
(ii) for any demand $a\in \calA$, at most one demand instance of $a$ is selected.
The profit of the solution is the sum of profits of the demand instance contained in it.
The goal is to find a feasible solution of maximum profit.

The communication among the processors is governed by the following rule:
two processors $P_1$ and $P_2$ are allowed to communicate, if
they have access to some common resource ($\Acc(P_1)\cap \Acc(P_2)\neq \emptyset$).

{\it Notation: }The following notation will be useful in our discussion.
Let $\calE$ denote the set of all edges over all the tree-networks;
any edge $e\in \calE$ is represented by a triple $\langle u, v, T\rangle$,
where $u$ and $v$ are vertices of $e$ and $T$ is the tree-network to which $e$ belongs.
For a tree-network $T$, let $\calD(T)$ denote the set of all demand instances belonging to $T$.
Any demand instance $d\in \calD(T)$ can be viewed as a path in $T$ and we denote this as $\mypath(d)$.
For a demand instance $d\in \calD(T)$ and an edge $e$ in $T$,
we say that $d$ is {\em active} on the edge $e$, if the $\mypath(d)$ includes $e$;
this is denoted $d\sim e$. We say that two demand instances $d_1$ and $d_2$ are
{\em overlapping}, if $d_1$ and $d_2$ belong to the same tree-network, 
and $\mypath(d_1)$ and $\mypath(d_2)$ share some edge;
the demands are said to non-overlapping, otherwise.
Two demand instances $d_1$ and $d_2$ are said to be {\em conflicting}, 
if both $d_1$ and $d_2$ belong to the same demand or they overlap;
otherwise, the demands are said to be non-conflicting.
We shall alternatively use the term {\em independent} to mean a pair of non-conflicting demands.
A set of demand instances $D$ is said to be {\em independent set},
if every pair of demand instances in $D$ is independent.
Notice that a feasible solution is nothing but an independent set of demand instances.

\section{LP and the Two-phase Framework}
Our algorithm uses the well-known primal-dual scheme and goes via a two-phase framework.
We first present the primal and the dual LPs and then discuss the framework.

\subsection{LP Formulation}
The LP and its dual are presented below.
For each demand instance $d\in \calD$, we introduce a primal variable $x(d)$.
The first set of primal constraints capture the fact that a feasible solution cannot
select two demand instances active on the same edge.
Similarly, the second set of primal constraints capture the fact that a feasible solution
can select at most one demand instance belonging to any demand.
For each demand $a\in \calA$ and each edge $e\in \calE$, the dual includes a variable $\alpha(a)$ and $\beta(e)$,
respectively. Similarly, for each demand instance $d\in \calD$, 
the dual includes a constraint; we call this the {\em dual constraint of $d$}.
Let $a_d$ denote the demand to which a demand instance $d$ belongs.
\begin{tabular}{p{3in}p{3in}}
\begin{eqnarray*}
\max \quad & & \sum_{d \in {\calD}} x(d) \cdot p(d) \\
\sum_{d\in \calD ~:~ d \sim  e} x(d) & \leq & 1 \quad  (\forall e\in \calE) \\
\sum_{d \in \Inst(a)} x(d) & \leq & 1 \quad (\forall a\in \calA) \\
x(d) & \geq & 0 \quad (\forall d\in \calD)
\end{eqnarray*}
&
\begin{eqnarray*}
\min \quad \sum_{a \in {\calA}} \alpha(a) &+& \sum_{e\in \calE} \beta(e) \\
\alpha(a_d) + \sum_{e~:~ d \sim  e} \beta(e) & \geq & p(d) \quad (\forall d\in \calD)\\
\alpha(a) & \geq & 0 \quad (\forall a\in \calA)\\
\beta(e) & \geq & 0 \quad (\forall e\in \calE)
\end{eqnarray*}
\end{tabular}



\subsection{Two-phase framework}
We formulate the ideas implicit in \cite{Pancj,Bar-Noy-Jacm,Lewin-Eytan} 
in the form of a two-phase framework, described next. Our algorithm would follow this framework.

{\it First Phase: }
The procedure initializes all the dual variables $\alpha(\cdot)$ and $\beta(\cdot)$ to $0$
and constructs an empty stack, and then it proceeds iteratively.
Consider an iteration. Let $U$ be the set of all demand instances whose dual constraints are still unsatisfied.
We select a suitable independent set $I\subseteq U$ (how to select $I$ is clarified below).
For each $d\in I$, we wish to increase (or {\em raise}) the value of the dual variables 
suitably so that the dual constraint of $d$ is satisfied tightly (i.e., the LHS becomes equal to the RHS).
For this purpose, we adopt the following strategy.
Consider each demand instance $d\in I$.
We first determine the {\em slackness} $s$ of the constraint, which is the difference between the 
LHS and RHS of the constraint: 
$s = p(d) - (\alpha(a_d) + \sum_{e~:~d\sim e} \beta(e))$.
We next select a suitable subset $\pi(d)$ consisting of edges on which $d$ is active
(how to select $\pi(d)$ is clarified below).
Next we compute the quantity $\delta(d)=s/(|\pi(d)|+1)$.
We then raise the value of $\alpha(a_d)$ by the amount $\delta(d)$;
and for each $e\in \pi(d)$, we raise dual variable $\beta(e)$ by the amount $\delta(d)$.
We see that the dual constraint is satisfied tightly in the process.
The edges $\pi(d)$ are called the {\em critical edges} of $d$.
We say that the demand instance $d$ is {\em raised} by the amount $\delta(d)$.
Finally, the independent set $I$ is pushed on to the stack (as a single object).
This completes an iteration.
In the above framework, in each iteration, we need to select
an independent set $I$ and the critical set of edges $\pi(d)$ for each $d\in I$. 
These are left as choices that must be made by the
specific algorithm constructed via this framework.
Similarly, the algorithm must also decide the termination condition for the first phase.

{\it Second Phase: } We consider the independent sets in the reverse order
and construct a solution $S$, as follows.
We initialize a set $D=\emptyset$ and proceed iteratively. 
In each iteration, the independent set $I$ on the top of the stack is popped.
For each $d\in I$, we add $d$ to $D$, if doing so does not violate feasibility 
(namely, $D\cup \{d\}$ is an independent set).
The second phase continues until the stack becomes empty.
Let $S=D$ be the feasible solution produced by the second phase.
This completes the description of the framework.

An important aspect of the above framework is that is parallelizable.
The set $I$ chosen in each iteration of the first phase is an independent set.
Hence, for any two demand instances $d_1,d_2\in I$, the LHS of the 
constraints of $d_1$ and $d_2$ do not share any dual variable.
Consequently, all the demand instances $d\in I$ can be raised simultaneously.

As we shall see, we can derive an approximation ratio for any 
algorithm built on the above framework, provided it satisfies the following condition,
which we call the {\em interference property}:
for any pair of overlapping demand instances $d_1$ and $d_2$ raised in the first phase,
if $d_1$ is raised before $d_2$, 
then $\mypath(d_2)$ must include at least one of the critical edges contained in $\pi(d_1)$. 

The following notation is useful in determining the approximation ratio.
Let $\xi\in [0,1]$ be any real number.
At any stage of the algorithm, we say that a demand instance $d\in \calD$ is
{\em $\xi$-satisfied}, if in the dual constraint of $d$,
the LHS is at least $\xi$ times the RHS:
$\alpha(a_d) + \sum_{e~:~d\sim e} \beta(e) \geq \xi \cdot p(d)$.
If the above condition is not true, then we say that $d$ is {\em $\xi$-unsatisfied}.

We shall measure the efficacy of an algorithm following the above framework using three parameters.
(1) {\em Critical set size $\Delta$: }Let $\Delta$ be the maximum cardinality of $\pi(d)$,
over all demand instances $d$ raised by the algorithm.
(2) {\em Slackness parameter $\lambda$: }Let $\lambda\in [0,1]$ be the largest number such that
at the end of the first phase, all the demand instances $d\in \calD$ are $\lambda$-satisfied.
(3) {\em Round complexity: }The number of iterations taken by the first phase.
The parameters $\Delta$ and $\lambda$ will determine the approximation ratio of the algorithm;
we would like to have $\Delta$ to be small and $\lambda$ to be close to $1$.
The round complexity determines the number of rounds taken by the algorithm when implemented in a distributed setting.
We say that the algorithm is {\em governed} by the parameters $\Delta$ and $\lambda$.

The following lemma provides an approximation guarantee for any algorithm satisfying the interference property.
The lemma is similar to Lemma 1 {in} the work of Panconesi and Sozio \cite{Pancc}. 
Let $\Opt$ denote the optimal solution to the input problem instance.

\begin{lemma}
\label{lem:AAA}
Consider any algorithm satisfying the interference property and governed by parameters $\Delta$ and $\lambda$.
Then the feasible solution $S$ produced by the algorithm satisfies 
$p(S)\geq \left(\frac{\lambda}{\Delta+1}\right)\cdot p(\Opt)$.
\end{lemma}
\proof
At the end of the first phase,
the algorithm produces dual variable assignments $\alpha(\cdot)$ and $\beta(\cdot)$.
Even though this assignment may not form a dual feasible solution,
it ensures that all the demand instances are $\lambda$-satisfied;
(intuitively, all the dual constraints are approximately satisfied).
It is easy to convert the assignment $\langle \alpha, \beta\rangle$ into a dual feasible solution
by scaling the values by an amount $1/\lambda$:
for each demand instance $d$, set $\wh{\alpha}(d) = \alpha(d)/\lambda$ and 
for each edge $e$, set $\wh{\beta}(e)=\beta(e)/\lambda$.
Notice that the $\langle \wh{\alpha},\wh{\beta}\rangle$ forms a feasible dual solution.

Let $\val(\alpha, \beta)$ and $\val(\wh{\alpha},\wh{\beta})$ be the objective
value of the dual assignment $\langle \alpha, \beta \rangle$
and the dual feasible solution $\langle \wh{\alpha},\wh{\beta}\rangle$,
respectively. By the weak duality theorem, $\val(\wh{\alpha},\wh{\beta}) \geq p(\Opt)$.
The scaling process implies that $\val(\wh{\alpha},\wh{\beta}) = \val(\alpha,\beta)/\lambda$.
We now establish a relationship between $\val(\alpha,\beta)$ and $p(S)$.

Let $R$ denote the set of all demand instances that are raised in the first phase.
The value $\val(\alpha,\beta)$ can be computed as follows.
For any $d\in R$, at most $\Delta+1$ dual variables are raised by an amount $\delta(d)$ (because $|\pi(d)|\leq \Delta$).
So, whenever a demand instance $d\in R$ is raised, the objective value raises at most $(\Delta+1)\delta(d)$.
Therefore,
\begin{eqnarray}
\nonumber
\val(\alpha,\beta) 
&=& \sum_{a\in \calA} \alpha(a) + \sum_{e\in \calE} \beta(e)\\
\label{eqn:AAA}
&\leq& (\Delta+1)\sum_{d\in R} \delta(d).
\end{eqnarray}

We next compute the profit of the solution $p(S$).
For a pair of demand instances $d_1, d_2\in R$, we say that $d_1$ is a {\em predecessor} of $d_2$,
if the pair is conflicting and $d_1$ is raised before $d_2$; in this case $d_2$ is said to be a {\em successor} of $d_1$.
For a demand instance $d\in R$, let $\mypred(d)$ and $\mysucc(d)$ denote the set of predecessors and successors of $d$,
respectively; we include $d$ in both the sets.

Consider an element $d\in S$. We claim that 
\begin{eqnarray}
\label{eqn:BBB}
p(d) \geq \sum_{d'\in \mypred(d)} \delta(d').
\end{eqnarray}
To see this claim, consider the iteration in which $d$ is raised.
At the beginning of this iteration the constraint of $d$ is unsatisfied 
and $d$ is raised to make the constraint tightly satisfied.
The interference property ensures the following condition:
any demand instance $d'\in \mypred(d)$ with $d'\neq d$ would have contributed a value of at least $\delta(d')$
to the LHS of the constraint (because the property enforces that the LHS includes at least one of raising dual variables
of $d'$). Thus, at the beginning of the iteration, the value of the LHS satisfies:
\[
p(d) \geq {\rm LHS} \geq \sum_{d'~:~d'\in \mypred(d) \mbox{~and~}d'\neq d} \delta(d').
\]
When $d$ is raised, LHS increases at least by $\delta(d)$ and the constraint becomes tight.
This proves the claim.

We can now compute a lowerbound on the profit of $S$:
\begin{eqnarray}
\nonumber
p(S) 
&=& \sum_{d\in S} p(d)\\
\nonumber
&\geq & \sum_{d\in S} \quad \sum_{d'\in \mypred(d)} \delta(d')\\
\nonumber
&=& \sum_{d'\in R} \quad \sum_{d\in \mysucc(d') \cap S} \delta(d')\\
\nonumber
&=& \sum_{d'\in R}\delta(d')\cdot |\mysucc(d')\cap S|.\\
\label{eqn:CCC}
&\geq & \sum_{d'\in R} \delta(d').
\end{eqnarray}
The second statement follows from (\ref{eqn:BBB}) and 
the last statement follows from the fact that for any $d'\in R$, either $d'$ belongs to $S$
or a successor of $d'$ belongs to $S$ (this is by the construction of the second phase).

Comparing (\ref{eqn:AAA}) and (\ref{eqn:CCC}), we see that $\val(\alpha,\beta)\leq (\Delta+1)p(S)$.
The lemma follows from the observations made at the beginning of the proof.
\qed

A local-ratio based sequential $3$-approximation algorithm for the unit height case of tree-networks
is implicit in the work of Lewin-Eytan \cite{Lewin-Eytan}.
This algorithm can be reformulated in the two-phase framework with parameters 
critical set size $\Delta=2$ and slackness $\lambda=1$ (however, the round complexity can be as high as $n$).
We present the above algorithm in Appendix \ref{sec:Lewin};
the purpose is to provide a concrete exposition of the two-phase framework.

Panconesi and Sozio \cite{Pancj} designed a distributed algorithm for the 
throughput maximization problem restricted to line-networks.
In terms of the two-phase framework, their algorithm satisfies the interference property
with critical set size $\Delta=3$ and slackness $\lambda=1/(5+\epsilon)$.
To this end, they partition the demand instances in to logarithmic number of groups
based on their lengths, wherein the lengths of any pair of demand instances found within the same group 
differ at most by a factor of 2. Then they exploit the property that if $d_1$ and $d_2$
are overlapping demand instances found within the same group , 
then $d_2$ is active either at the left end-point, the right end-point or the mid-point of $d_1$.
This way, they satisfy the interference property with $\Delta=3$.
We do not know how to extend such a length-based ordering to our setting of tree-networks.
Consequently, designing an ordering satisfying the interference property with a constant $\Delta$ 
turns out to be more challenging. Nevertheless, we show an ordering for which $\Delta=6$.
Furthermore, we shall present a method for improving the slackness parameter $\lambda$ to $(1-\epsilon)$.
The notion of {\em tree-decompositions} and {\em layered decompositions} form the core components
of our algorithms.

\section{Tree-Decompositions and Layered Decompositions}
We first define the notion of tree-decompositions and show how to
construct tree decompositions with good parameters. 
Then, we show how to transform tree decompositions into layered decompositions.

Let $H$ be a rooted tree defined over the vertex-set $V$ with $g$ as the root.
For a node $x$, define its {\em depth} to be the number
of nodes along the path from $g$ to $x$; the root $g$ itself is defined to have to depth $1$.
With respect to $H$, a node $y$ is said to be an {\em ancestor} of $x$,
if $y$ appears along the path from $g$ to $x$; in this case, $x$ is said to be a {\em descendent} of $y$.
By convention, we do not consider $x$ to be an ancestor or descendent of itself.
For a node $z$ in $H$, let $C(z)$ be the set consisting of $z$ and its descendents in $H$.

\subsection{Tree-decomposition: Definition}
Let $T\in \calT$ be a tree-network defined over the input vertex-set $V$ consisting of $n$ vertices.
A subset of nodes $C\subseteq V$ is called a {\em component}, if $C$ induces a (connected) subtree in $T$.
We say that a node $x\in V-C$ is a {\em neighbor} of $C$, if $x$ is adjacent to some node in $C$.
Let $\Gamma[C]$ denote the set of neighbors (or {\em neighborhood}) of $C$.
Notice that for any two nodes $x\in C$ and $y\not\in C$, the path between $x$ and $y$ must pass through 
some node in the neighborhood $\Gamma[C]$.

Let $T$ be a tree-network and $H$ be a rooted-tree defined over $V$ with $g$ as the root.
We say that $H$ is a {\em tree decomposition} for $T$, 
if the following conditions are satisfied: 
(i) for any demand instance $d\in \calD(T)$, if $d$ passes through nodes $x$ and $y$ then $d$ also passes through
$\LCA(x,y)$, which is the least common ancestor of $x$ and $y$ in $H$;
(ii) for any node $z$ in $H$, $C(z)$ forms a component in $T$.

For a node $z\in H$, let $\chi(z)$ denote the set of neighbors of the component $C(z)$, i.e., $\Gamma[C(z)]$.
We call $\chi(z)$ the {\em pivot set} of $z$.
Clearly, for any nodes $x\in C(z)$ and $y\not\in C(z)$, the path between $x$ and $y$ in $T$ must 
pass through one of the nodes in $\chi(z)$.
We shall measure the efficacy of a tree decomposition $H$ using two parameters:
(i) {\em pivot size $\theta$: } this is the maximum cardinality of $\chi(z)$ over all $z\in V$;
(ii) the depth of the tree. 

\begin{figure}
\centering
\includegraphics[width=4.00in]{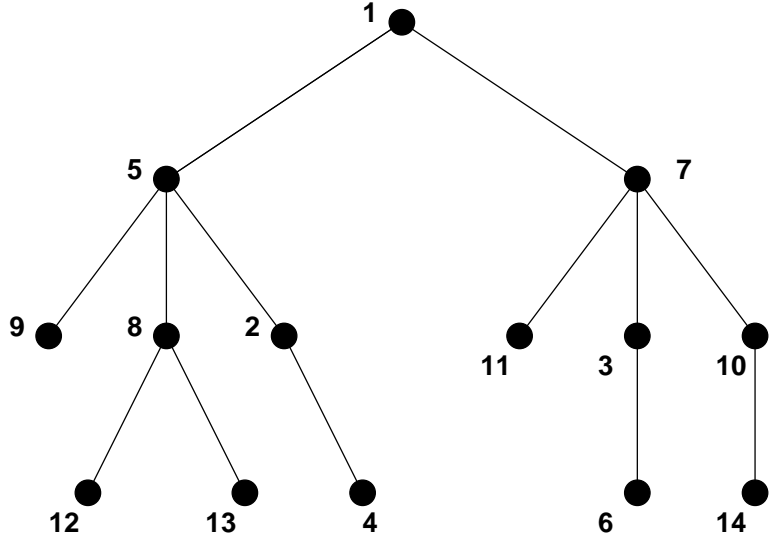}
\caption{
Tree-decompositions: Illustration.
}
\label{fig:6}
\end{figure}

See Figure \ref{fig:6} for an illustration.
This figure shows an example tree-decomposition for the tree-network shown in Figure \ref{fig:5}.
The demand instance $\pair{4}{13}$ passes through nodes $2$ and $8$; it also
passes through $\LCA(2,8)=5$. For the node $2$, the component $C(2)=\{2,4\}$; its pivot set is $\chi(2)=\{1,5\}$.
On the other hand, $C(5)=\{5,9,8,2,12,13,4\}$ and its pivot set is $\chi(5)=\{1\}$.
This tree-decomposition has depth $4$ and pivot set size $\theta=2$.

We note that it is not difficult to design tree-decompositions with 
parameters $\pair{\depth=n}{~\theta=1}$ or $\pair{\depth=\log n}{~\theta=\log n}$.
As it turns out the depth of the tree-decomposition will determine the number of rounds,
whereas the pivot size $\theta$ will determine the approximation ratio.
Thus, neither of these two tree-decompositions would yield an algorithm
that runs in polylogarithmic number of rounds, while achieving a constant factor approximation ratio.
Our main contribution is a tree-decomposition with parameters $\pair{\depth=2\log n}{~\theta=2}$
(we call this the {\em ideal tree-decomposition}).
Interestingly, the ideal tree-decomposition builds on the two simpler tree-decompositions mentioned above.
For the sake of completeness, the two simpler tree-decompositions are discussed next.

\subsection{Two Simple Tree-decompositions}
\label{sec:simple-trees}
Here, we present two tree decompositions called {\em root-fixing tree decomposition}
and {\em balancing tree decomposition}. The first decomposition has pivot size $\theta=1$,
but its depth can be as high as $n$. The second decomposition has depth $\log n$,
but its pivot size $\theta$ can be as high as $\log n$.
\\

\noindent
{\bf Root-fixing Decomposition: } 
Let $T\in \calT$ be any input tree-network. Convert $T$ into a rooted-tree by 
arbitrarily picking a node $g\in V$ as the root; let the resulting rooted-tree be $H$.
It is easy to see that $H$ is a tree decomposition for $T$.
Consider any node $z$ and let $z'$ be its parent in $H$; let $C(z)$ be the descendants of $z$ including $z$ itself.
Notice that for any $x\in C(z)$ and $y\not\in C(z)$, the path between $x$ and $z$ must pass through
the parent $z'$. Thus the component $C(z)$ has only one neighbor.
We see that $H$ has pivot size $\theta=1$; however, the depth of $H$ can be as high as $n$.
Figure \ref{fig:5} shows a root-fixing decomposition; the chosen root is node $1$.
The sequential algorithm given in Section \ref{sec:Lewin} implicitly uses the root-fixing tree decomposition.
\\

\noindent
{\bf Balancing tree decomposition: }
Let $T\in \calT$ be a tree-network. Consider a component $C\subseteq V$
and let $T(C)$ be the (connected) subtree induced by $C$.
Let $z$ be a node in $C$. If we delete the node $z$ from $T(C)$,
the tree $T(C)$ splits into subtrees $T_1, T_2, \ldots, T_s$ (for some $s$).
Let $C_1, C_2, \ldots, C_s$ be the vertex-set of these subtrees.
Every node in $C-\{z\}$ is found in some component $C_i$.
We say that the node $z$ {\em splits} $C$ into components $C_1, C_2, \ldots C_s$.
The node $z$ is said to be a {\em balancer} for $C$, if for all $1\leq i\leq s$, $|C_i\leq \floor{|C|/2}$.
The following observation is easy to prove:
any component $C\subseteq V$ contains a balancer $z$.

Our procedure for constructing the tree decomposition for $T$ works recursively
by calling a procedure {\BuildBTD} ({\em build balanced tree decomposition}).
The procedure takes as input a component $C\subseteq V$ and outputs a rooted-tree
having $C$ as the vertex-set. It works as follows.
Given a component $C$, find a balancer $z$ for $C$.
Then split $C$ by $z$ and obtain components $C_1, C_2, \ldots, C_s$ (for some $s$).
Each component $C_i$ has size at most $\floor{C/2}$.
For $1\leq i\leq s$, call the procedure {\BuildBTD} recursively on the component $C_i$
and obtain a tree $H_i$ with $g_i$ as the root. Construct a tree $H$ by making $z$ as the root
and $g_1, g_2, \ldots, g_s$ as its children. Return the tree $H$.

Given a tree-network $T$, we obtain a rooted-tree $H$ by calling {\BuildBTD} with the whole
vertex-set $V$ as the input. 
It is easy to see that for any node $z$, $C(z)$ forms a component in $T$.
For any node $z$ in $H$ with children $z_1, z_2, \ldots, z_s$ (for some $s$),
$C(z_1), C(z_2), \ldots, C(z_s)$ are nothing but the components obtained by splitting $C(z)$ by $z$.
This implies that $H$ satisfies the first property of tree decompositions.
Since the size of the input component drops by a factor of two in each iteration,
the depth of $H$ is at most $\ceil{\log n}$.
Consider any node $z$ in $H$ and let $C(z)$ be the set consisting of descendants of $z$ and $z$ itself.
Observe that for any node $x\in C(z)$ and $y\not\in C(z)$, the path between $x$ and $y$
must pass through one of the ancestors of $z$ in $H$ (because of the first property of tree decompositions).
In other words, the neighborhood of $C(z)$ is contained within the set of ancestors of $z$.
The number of ancestors is at most $\ceil{\log n}$ and
hence, the pivot size of $H$ is at most $O(\log n)$.
Figure \ref{fig:6} shows an example balancing tree-decomposition for the tree given in Figure \ref{fig:5}.

\subsection{Ideal Tree-decomposition}
In this section, we present the ideal tree-decomposition with parameters $\pair{\depth=2\log n}{~\theta=2}$.
The ideal tree-decomposition also goes via the notion of balancers.
Recall that any component $C\subseteq V$ contains a balancer $z$.

Fix a tree-network $T$ and we shall construct an ideal tree decomposition $H$ for $T$ with pivot set size $\theta=2$
and depth $O(\log n)$. Intuitively, the tree $H$ will be constructed recursively.
In each level of the recursion, we will add two nodes to the tree: a balancer and a node that we call a {\em junction}.
The output tree-decomposition will have depth at most $2\ceil{\log n}$.

The construction works via a recursive procedure {\BuildITD} ({\em build ideal tree decomposition}).
The procedure {\BuildITD} takes as input a set $C\subseteq V$ forming a component in $T$.
As a precondition, it requires the component $C$ to satisfy the important property that $C$ has at most
two neighbors in $T$.
It outputs a rooted-tree $H$ with $C$ as the vertex set
having depth at most $2\ceil{\log |C|}$ such that for any node $x\in C$, 
the number of neighbors of $C(x)$ is at most $2$,
where $C(x)$ is the set consisting of $x$ and its descendants in $H$.

The procedure {\BuildITD} works as follows. We first find a balancer $z$ for the component $C$.
The node $z$ splits $C$ into components $C_1, C_2, \ldots, C_s$. 
We shall consider two cases based on whether $C$ has a single neighbor or two neighbors.

\begin{figure}[t]
\centering
\includegraphics[width=6in]{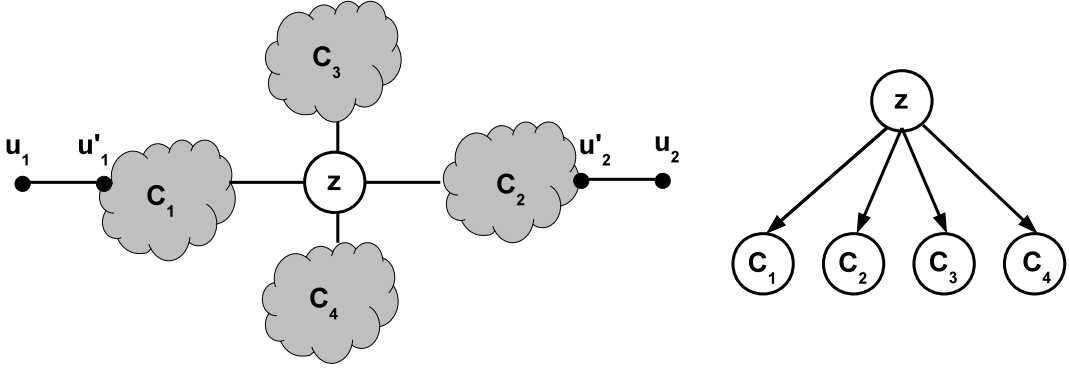}
\caption{Illustration for ideal tree-decomposition.}
\label{fig:3}
\end{figure}

{\it Case 1: }This is the easier case where $C$ has only one neighbor, say $u_1$.
See Figure \ref{fig:3}. For this case, ignore the nodes $u_2$ and $u_2'$.
Let $u_1'$ be the node in $C$ which is adjacent to $u_1$ and 
without loss of generality, assume that $C_1$ is the component to which $u_1'$ belongs.
Observe that $\Gamma(C_1)=\{u_1,z\}$ and for all $i\geq 2$, $\Gamma(C_i)=\{z\}$.
In other words, all the components $C_i$ have at most two neighbors.
That is, they all satisfy the precondition set by the procedure.
For each $1\leq i\leq s$, we recursively call the procedure {\BuildITD} on the component $C_i$
and obtain a tree $H_i$ with $g_i$ as the root. 
We construct a tree $H$ by making $z$ as the root and $g_1, g_2,\ldots, g_s$ as its children. 
Then, the rooted-tree $H$ is returned. 

{\it Case 2: }Now consider the case where $C$ has two neighbors, say $u_1$ and $u_2$.
Let $u_1'$ and $u_2'$ be the nodes in $C$ which are neighbors of $u_1$ and $u_2$, respectively.
We consider two subcases.

{\it Case 2(a): }
The first subcase is when $u_1'$ and $u_2'$ lie in two different components, say $C_1$ and $C_2$, respectively. 
See Figure \ref{fig:3}.
Observe that $\Gamma(C_1)=\{u_1,z\}$, $\Gamma(C_2)=\{u_2,z\}$ and for all $i \geq 3$,
$\Gamma(C_i)=\{z\}$. Hence all the components $C_i$ satisfy the precondition set by the procedure.
For each $1\leq i\leq s$, we call the procedure {\BuildITD} with $C_i$ as input and obtain 
a tree $H_i$. We construct a tree $H$ by making the balancer $z$ as the root and $g_1, g_2,\ldots, g_s$ as its children. 
Then, the rooted-tree $H$ is returned. 

\begin{figure}[t]
\centering
\includegraphics[width=6in]{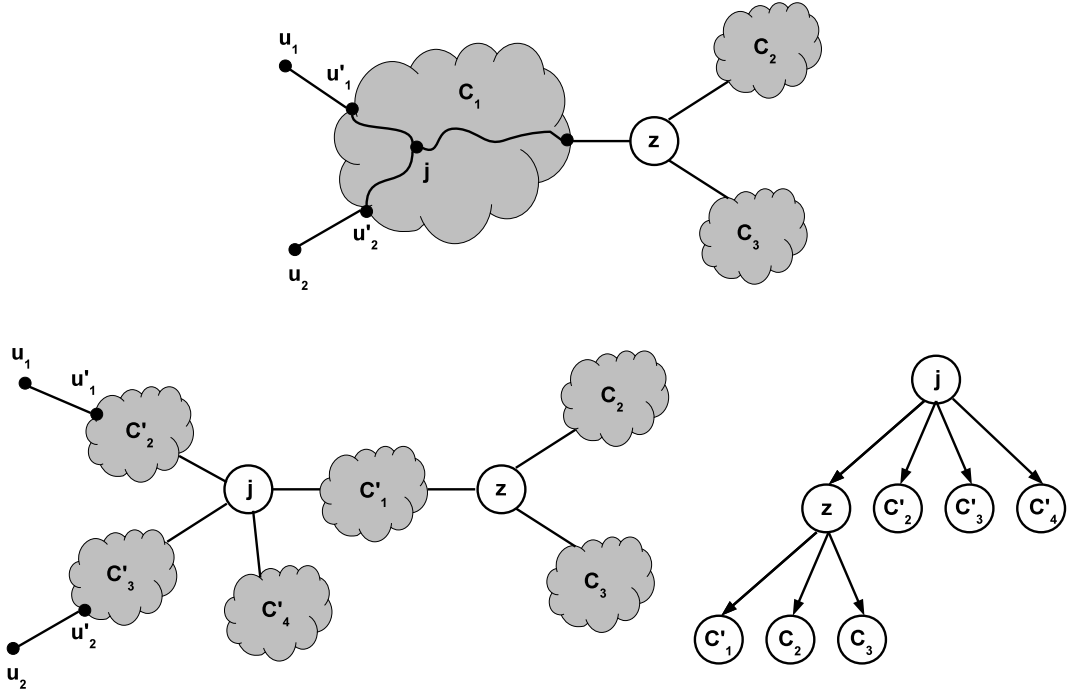}
\caption{Illustration for Case 2(b) of ideal tree-decomposition}
\label{fig:4}
\end{figure}

{\it Case 2(b): }
Now consider the second and comparatively more involved subcase wherein $u_1'$ and $u_2'$ belong to 
the same component, say $C_1$. See Figure \ref{fig:4}.
Observe that there exists a unique node $j\in C_1$ such that all the three paths 
$u_1\leadsto u_2$, $u_1\leadsto z$, and $u_2\leadsto z$ pass through $j$. We call $j$ as the {\em junction}.
Spilt the component $C_1$ by the node $j$ to obtain components $C_1', C_2', \ldots, C'_{s'}$ (for some $s'$).
Observe that among $C_1',C_2',\ldots, C'_{s'}$, 
there exists three distinct components such that $z$ is a neighbor of the first component,
and $u_1'$ and $u_2'$ belong to the other two components;
without loss of generality, let these components be $C'_1$, $C_2'$ and $C'_3$, respectively.
We see that for $2\leq i\leq s$, $\Gamma(C_i)=\{z\}$; moreover,
$\Gamma(C_1') = \{j,z\}$, $\Gamma(C_2')=\{u_1,j\}$, $\Gamma(C_3')=\{u_2,j\}$
and for $4\leq i\leq s'$, $\Gamma(C_i') = \{j\}$.
Thus, all the components $C_2,C_3,\ldots,C_s$ and $C_1',C_2',\ldots, C_{s'}'$
satisfy the precondition set by the procedure.
For each $2\leq i\leq s$, we call the procedure {\BuildITD} recursively with
$C_i$ as input and obtain a tree $H_i$ with $g_i$ as the root.
For each $1\leq i\leq s'$, we call the procedure {\BuildITD} recursively with
$C_i'$ as input and obtain a tree $H_i'$ with $g_i'$ as the root.
Construct a tree $H$ as follows. Make the junction $j$ as the root;
make $g_2',g_3',\ldots, g_{s'}'$ as the children of $j$;
make $z$ as a child of $j$; make $g_1'$ and $g_2, g_3,\ldots,g_s$ as the children of $z$.
Return the rooted-tree $H$. This completes the description of the procedure {\BuildITD}.

By induction, we can argue that {\BuildITD} satisfies the intended property:
for any node $x\in C$, the number of neighbors of $C(x)$ is at most $2$.
As an example, consider the subcase in which $u_1'$ and $u_2'$ belong to the same component 
(the case where a junction $j$ is created). The procedure creates only two nodes $j$ and $z$ on its own
and the rest of the nodes in $H$ are created by the recursive calls. 
Consider the node $j$.
It is guaranteed that the input component $C$ has at most two neighbors (this is the precondition
set by the procedure). Since $C(j)=C$, we see that $j$ satisfies the property.
Now, consider the node $z$. The component $C(z)$ is the union 
of $C_2, C_3,\ldots, C_s$ and $C_1'$. We have that $\Gamma[C(z)]=\{j\}$.
Thus, $z$ also satisfies the property. The rest of the nodes satisfy the property by induction.

Let us now analyze the depth of the tree $H$ output by the procedure.
Since $z$ is a balancer for $C$, the components $C_1, C_2, \ldots, C_s$ have size at most $\floor{C/2}$.
Moreover, since $C_1', C_2',\ldots, C_{s'}'$ are subsets of $C_1$,
these components also have size at most $\floor{C/2}$.
Thus, all the components input to the recursive calls have size at most $\floor{C/2}$. 
Thus, by induction, $H$ has depth at most $2\lceil \log C\rceil$.

We next show how to construct a tree decomposition $H$ for the tree-network $T$.
First, find a balancer $g$ for the entire vertex-set $V$ and split $V$ into components
$C_1, C_2, \ldots, C_s$. For each component $C_i$, $\Gamma[C_i]=\{g\}$.
For each $1\leq i\leq s$, call the procedure $\BuildITD$
with $C_i$ as input and obtain a tree $H_i$ with $g_i$ as the root.
Construct a tree $H$ by making $g$ as the root and each $g_i$ as its children.
Return $H$.

We can argue that for any node $z$ in $H$, $C(z)$ forms a component in $T$.
Furthermore, for any node $z$ in $H$ with children $z_1, z_2, \ldots, z_s$ (for some $s$),
$C(z_1), C(z_2), \ldots, C(z_s)$ are nothing but the components obtained by splitting $C(z)$ by $z$.
This implies that $H$ satisfies the first property of tree decompositions.
It follows that $H$ is indeed a tree decomposition.
The depth of $H$ is at most $2\lceil \log n\rceil$. The properties of the {\BuildITD} procedure
ensure that the pivot size of $H$ is at most $2$. We have the following result

\begin{lemma}
\label{lem:BBB}
For any tree-network $T\in \calT$, there exists a tree decomposition $H$ (called the ideal tree decomposition)
with depth $O(\log n)$ and pivot size $\theta=2$.
\end{lemma}

\subsection{Layered Decompositions}
In this section, we define the notion of {\em layered decompositions} and 
show how to transform tree decompositions into layered decompositions.

Let $T\in \calT$ be a tree-network. A layered decomposition of $T$ is a pair 
$\sigma$ and $\pi$, where $\sigma$ is a partitioning of $\calD(T)$ into a
sequence of groups $G_1, G_2, \ldots, G_{\ell}$ and $\pi$ maps each demand instance $d\in \calD(T)$
to a subset of edges in $\mypath(d)$. The following property should be satisfied:
for any $1\leq i\leq j\leq \ell$ and for any pair of demand instances $d_1\in G_i$ and $d_2\in G_j$,
if $d_1$ and $d_2$ are overlapping, then $\mypath(d_2)$ should include at least one of the 
edges in $\pi(d_1)$. The edges in $\pi(d)$ are called the {\em critical edges} of $d$.
The value $\ell$ is called the {\em length} (or depth) of the decomposition.

Notice that similarity between the inference property and the notion of layered decompositions.
We shall measure the efficacy of a layered decomposition by two parameters:
(i) {\it Critical set size $\Delta$} -  this is the maximum cardinality of $\pi(d)$ over all 
demand instances $d\in \calD(T)$; (ii) the length $\ell$ of the sequence.
Our goal is to construct a layered decomposition with length $O(\log n)$
and critical set size $\Delta=6$. Towards that goal we shall show how to transform 
tree-decompositions into layered decompositions. The following notations are useful for this purpose.

\begin{figure}
\centering
\includegraphics[width=4.00in]{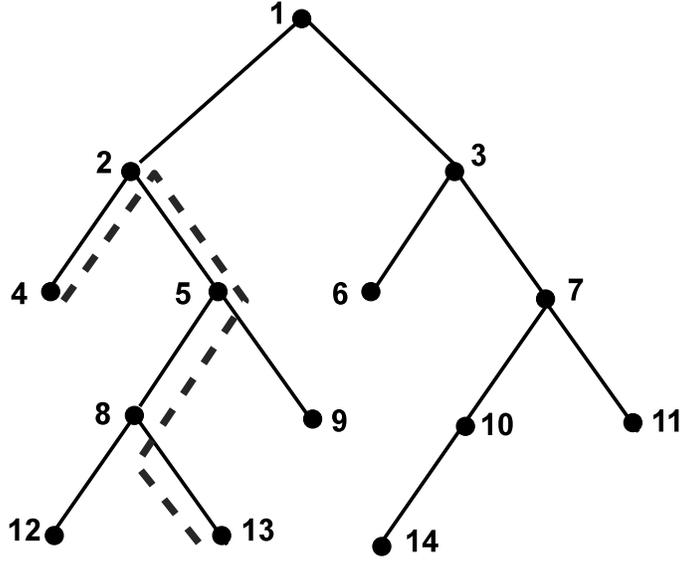}
\caption{
An example tree-network. 
}
\label{fig:5}
\end{figure}

Let $T\in \calT$ be tree-network and $H$ be a tree-decomposition for $T$
with pivot size $\theta$ and depth $\ell$. 
For a demand instance $d$, let $\mu(d)$
be the node with the least depth in $H$ among all the nodes that $\mypath(d)$ passes through.
The first property of tree decompositions ensure that $\mu(d)$ is unique.
We say that $d$ is {\em captured} at $\mu(d)$. See Figure \ref{fig:6}.
In this figure, the demand $\pair{4}{13}$ is captured at node $5$.
Let $d\in \calD(T)$ be a demand instance and $u$ be a node in $T$.
Observe that there exists a unique node $y$ belonging to $\mypath(d)$
such that the path from $u$ to $y$ does not pass through any other node in $\mypath(d)$.
We call $y$ as the {\em bending point} of $d$ with respect to $u$.
For a node $y$ in $\mypath(d)$, we call the edges on $\mypath(d)$ adjacent to $y$ as the {\em wings} of $y$ on $\mypath(d)$.
If $y$ is an end-point of $d$, there will be only one wing; otherwise, there will be two wings.
See Figure \ref{fig:5}.
In this figure, with respect to nodes $3$ and $9$, 
the bending points of the demand $d=\pair{4}{13}$ are $2$ and $5$, respectively.
With respect to $\mypath(d)$, node $4$ has only one wing $\pair{4}{2}$, while node
$8$ has two wings $\pair{5}{8}$ and $\pair{8}{13}$.

Lemma \ref{lem:CCC} shows how to transform a tree-decomposition into a layered decomposition.
The lemma is proved by categorizing the demand instances into groups
$G_1, G_2, \ldots, G_{\ell}$, where $G_i$ consists of all demand instances
captured at a node with depth $\ell-i+1$.
For a demand instance $d\in \calD(T)$, let $z=\mu(d)$.
The set $\pi(d)$ is constructed as follows:
(i) we include the wings of $z$ on $\mypath(d)$;
(ii) for each neighbor $u$ of $C(z)$, taking $y$ to be the bending point of $\mypath(d)$
with respect to $u$, we include the wings of $y$ on $\mypath(d)$.

\begin{lemma}
\label{lem:CCC}
Let $T\in \calT$ be a tree-network and $H$ be a tree decomposition for $T$
with pivot size $\theta$ and depth $\ell$. Then $H$ can be transformed into a layered
decomposition $\pair{\sigma}{\pi}$ with critical set size $\Delta=2(\theta+1)$
and length $\ell$.
\end{lemma}
\proof
For $1\leq i\leq \ell$, let $\wt{G}_i$ to be the set consisting of all demand instances 
$d$ such that depth of $\mu(d)$ is $i$. 
We define $\sigma$ to be the reverse of $\wt{G}_1, \wt{G}_2, \ldots, \wt{G}_{\ell}$;
namely, let $\sigma=G_1, G_2, \ldots, G_{\ell}$, where $G_i = \wt{G}_{\ell-i+1}$, for $1\leq i\leq \ell$.
Thus, in $\sigma$, the demand instances captured at the nodes having the highest depth are placed in $G_1$
and the demand instances captured at the root are placed in $G_{\ell}$.
We now show how to construct the critical set $\pi(d)$ for each demand instance $d\in \calD(T)$.
Let $z=\mu(d)$ be the node in $H$ where $d$ is captured.
Add the wing(s) of $z$ on $\mypath(d)$ to $\pi(d)$.
Then, consider the component $C(z)$ consisting of $z$ and its descendents in $H$.
Let $U=\{u_1, u_2, \ldots, u_s\}$ be the neighbors of $C(z)$, where $s\leq \theta$.
For $1\leq i\leq s$, let $y_i$ be the bending point of $d$ with respect to $u_i$;
add the wing(s) of $y_i$ on $\mypath(d)$ to $\pi(d)$. 
Notice that $\pi(d)$ has at most $2(\theta+1)$ edges.
This completes the construction of $\sigma$ and $\pi(\cdot)$.

We now argue that the construction satisfies the properties of layered decompositions.
Consider any two groups $G_i$ and $G_j$ such that $i\leq j$.
Consider two overlapping demand instances $d_1\in G_i$ and $d_2\in G_j$. 
Let $z_1=\mu(d_1)$ and $z_2=\mu(d_2)$ be the nodes in $H$ where $d_1$ and $d_2$ are captured, respectively.
We consider two cases: (1) $z_2 \in C(z_1)$; (2) $z_2\not\in C(z_1)$. 

{\it Case 1: } In this case, $z_2$ must be the same as $z_1$ 
(otherwise, we have $\depth(z_2) > \depth(z_1)$; this would contradict $i\leq j$).
Therefore, $\mypath(d_2)$ should include at least one of the wings of $z_1$ on $\mypath(d_1)$. 
Recall that the wing(s) of $z_1$ on $\mypath(d_1)$ are included in $\pi(d_1)$.

{\it Case 2: }
By the LCA property of tree-decompositions, $\mypath(d_1)$ will be constained within the component $C(z_1)$.
We have that $\mypath(d_2)$ goes through the node $z_2$ found outside of $C(z_1)$;
moreover, it also goes through some node found within $C(z_1)$ (since $d_1$ and $d_2$ overlap).
By the second property of tree decompositions, such a path must also 
pass through one of the neighbors of $C(z_1)$; let $u$ be such a neighbor.
Let the bending point of $\mypath(d_1)$ with respect to $u$ be $y$.
Since $\mypath(d_2)$ passes through $u$ and overlaps with $\mypath(d_1)$,
the $\mypath(d_2)$ must also pass through the bending point $y$. 
It follows that $\mypath(d_2)$ must include one of the wings of $y$ on $\mypath(d_1)$.
Recall that the wing(s) of $y$ on $\mypath(d_1)$ are included in $\pi(d_1)$.
\qed

By applying Lemma \ref{lem:CCC} for the ideal tree decomposition (given by Lemma \ref{lem:BBB}),
we establish the following result.

\begin{lemma}
\label{lem:DDD}
For any tree-network $T\in \calT$, we can construct a layered decomposition with 
critical set size $\Delta = 6$ and length at most $O(\log n)$.
\end{lemma}

\begin{figure}[t!]
\begin{center}
\begin{boxedminipage}{\hsize}
\begin{tabbing}
xx\=xx\=xx\=xx\=xxx\=xxx\=\kill
\textbf{Begin}  \\
\> // Initialize \\
\> For all $a\in \calA$, set $\alpha(a)=0$; for all $e\in \calE$, set $\beta(e)=0$.\\
\> Initialize an empty stack.\\
\> Let the input set of tree-networks be $\calT= \{T_1, T_2, \ldots, T_r\}$.\\
\> For each tree-network $T_q$\\
\> \> Invoke Lemma \ref{lem:DDD} on $T_q$ and \\
\> \> \> obtain a layered decomposition $\sigma_q=G_1^{(q)}, G_2^{(q)}, \ldots, G_{\ell_q}^{(q)}$ and a mapping $\pi_q$.\\
\> \> Let $\ell_{\max} = \max_{q} \ell_q$. \\
\> \> For each $d\in \calD(T_q)$, define $\pi(d)=\pi_q(d)$.\\
\> For each $k$ = $1$ to $\ell_{\max}$, \\
\> \> Define $G_k = \cup_{q=1}^r G^{(q)}_k$.\\
\>\\
\> // First phase\\
\> For $k=1$ to $\ell_{\max}$ \quad\quad//Epochs\\
\> \> Let $b$ be the smallest integer such that $(14/15)^b\leq \epsilon$ \\
\> \> For $j=1$ to $b$ \quad \quad //Stages \\
\> \> \> While \quad \quad // Steps or iterations.\\
\> \> \> \> Let $U=\{d\in G_k~:~\mbox{$d$ is $(1-(14/15)^j)$-unsatisfied}\}$\\
\> \> \> \> If $U=\emptyset$, exit the loop.\\
\> \> \> \> Find a maximal independent set $I$ contained within $U$\\
\> \> \> \> For each $d\in I$\\
\> \> \> \> \> Compute slackness: $s=p(d) - \alpha(a_d) - \sum_{e:d\sim e} \beta(e)$.\\
\> \> \> \> \> Compute: $\delta(d) = s/(|\pi(d)|+1)$.\\
\> \> \> \> \> Raise the variables: $\alpha(a_d)\leftarrow \alpha(a_d)+\delta(d)$; 
		for all $e\in \pi(d)$, $\beta(e) \leftarrow \beta(e)+\delta(d)$.\\
\> \> \> \> Push $I$ into the stack (as a single object).\\
\>\\
\> // Second Phase\\
\> \> $S=\emptyset$.\\
\> \> While(stack not empty)\\
\> \> \> Pop the top element $I$ of the stack\\
\> \> \> For each $d\in I$\\
\> \> \> \> If $S\cup \{d\}$ is an independent set, then add $d$ to $S$.\\
\> Output $S$.\\
\textbf{End}
\end{tabbing}
\end{boxedminipage}
\end{center}
\caption{Pseudocode of the overall algorithm}
\label{fig:full-pseudo}
\end{figure}

\section{Distributed Algorithm}
\label{sec:distributed}
In this section, we prove the main result of the paper by exhibiting
a two-phase procedure with critical set size 
$\Delta=6$ and slackness parameter $\lambda=(1-\epsilon)$, for any constant $\epsilon>0$.

Let the input tree networks be $\calT=\{T_1, T_2, \ldots, T_r\}$.
For each tree-network $T_q$, invoke Lemma \ref{lem:DDD} and obtain a layered decomposition
$\sigma_q=G_1^{(q)}, G_2^{(q)}, \ldots, G_{\ell_q}^{(q)}$ of length $\ell_q$ and a mapping $\pi_q$.
Let $\ell_{\max} = \max_{q} \ell_q$. The lemma guarantees that $\ell_{\max}$ is $O(\log n)$
and all the critical set sizes are at most $\Delta=6$. 
Let $\Delta'=\Delta+1$ and $\xi=(2\Delta')/(2\Delta'+1)=14/15$.
For the ease of exposition,
we combine all the mapping functions into single mapping function $\pi$, as follows.
For each tree-network $T_q$ and demand instance $d\in \calD(T_q)$, define $\pi(d)=\pi_q(d)$.

For each $1\leq k\leq \ell_{\max}$, let $G_k$ be union of the $k$th components
of all the layered decompositions: $G_k = \cup_{q=1}^r G^{(q)}_k$.
The algorithm would follow the two-phase framework.
All the dual variables are initialized to zero and an empty stack is created.
The first phase is split into $\ell_{\max}$ epochs.
Epoch $k$ will process the group $G_k$. Our goal is to ensure that at the end of the epoch,
all the demand instances in $G_k$ are $(1-\epsilon)$-satisfied.
Each epoch is divided into multiple stages, with each stage making a gradual progress towards the goal.
We will ensure that at the end of stage $j$, all the demand instances in $G_k$ are $(1-\xi^j)$-satisfied.
Each stage is split into multiple steps (each step corresponds to an iteration of the two-phase framework).
A typical step is explained next.
Let $U$ be the set of all demand instances in $G_k$ that are $(1-\xi^j)$-unsatisfied.
Find a maximal independent set $I$ contained within $U$.
For all demand instances $d\in I$, raise the demand instance $d$ as prescribed by the framework,
taking $\pi(d)$ to be the critical edges.
Namely, for all demand instances $d\in I$, perform the raising as follows.
Compute the slackness $s=p(d)-\alpha(a_d)-\sum_{e~:~d\sim e} \beta(e)$
and $\delta(d) = s/(|\pi(d)|+1)$. Raise the dual variable $\alpha(a_d)$ by the amount $\delta(d)$
and for all $e\in \pi(d)$, raise the dual variable $\beta(e)$ by $\delta(d)$.
The stage is completed when all the demand instances in $G_k$ are $(1-\xi^j)$-satisfied
and we proceed to the next stage. The epoch is completed when all the demand instances in $G_k$
are $(1-\epsilon)$-satisfied. The second phase is the same as that of the two-phase framework.
The pseudocode is provided in Figure \ref{fig:full-pseudo}.

Let us analyze the number of steps (or iterations) taken by the above algorithm.
The number of epochs is $\ell_{\max}$, which is $O(\log n)$.
Each epoch has at most $\log_{\xi}{\epsilon} = O(\log(1/\epsilon))$ stages.
The lemma below provides a bound on the number of steps taken by each stage.
It follows that the total number of communication rounds
is at most $O(\TMIS\log n\log(1/\epsilon)\log(p_{\max}/p_{\min}))$,
where $\TMIS$ is the number of (communication) rounds needed to find a maximal independent set (see Introduction).

\begin{lemma}
\label{lem:EEE}
Consider any epoch $k$ and stage $j$ within the epoch.
The number of steps taken by the stage is at most $O(\log(p_{\max}/p_{\min}))$.
\end{lemma}
\proof
Let the number of steps taken by the stage be $L$.
For $1\leq i\leq L$, let $U_i$ be the demand instances in $G_k$ that
are $(1-\xi^j)$-unsatisfied at the beginning of step $i$.
Let $I_1, I_2, \ldots, I_L$ be the sequence of maximal independent sets computed in these steps.
For two demand instances $d_1, d_2\in G_k$, we say that $d_1$ {\em kills} $d_2$ in step $i$,
if $d_1\in I_i$, $d_2\in U_{i+1}$, and $d_1$ and $d_2$ are conflicting.
Intuitively, both $d_1$ and $d_2$ are present in $U_i$, and both are contenders for the maximal independent $I_i$.
Of the two, $d_1$ got selected in $I_i$ and $d_2$ was omitted;
even after the demand instances in $I_i$ were raised, $d_2$ was still $(1-\xi^j)$-unsatisfied.
Since $d_1$ and $d_2$ are conflicting, only one of them can be included in the independent set. 
We imagine that $d_1$ ``kills" $d_2$.

\begin{claim}
Suppose $d_1$ kills $d_2$ in step $i$. Then, their profits satisfy $p(d_2)\geq 2p(d_1)$
\end{claim}

We now prove the claim. Since $d_1\in I_i$, the demand instance is $(1-\xi^j)$-unsatisfied
at the beginning of step $i$. Hence, the difference between the LHS and RHS 
of the constraint is at least $\xi^j\cdot p(d_1)$. The number dual variables raised for $d_1$ is at most $\Delta+1$.
Hence, 
\[
\delta(d_1)\geq \frac{\xi^j\cdot p(d_1)}{(\Delta+1)}
\]
Since $d_1$ and $d_2$ are conflicting, either it is the case that $d_1$ and $d_2$ belong to the same demand $a$
or they belong to the same tree-network $T_q$ (for some $q$) and overlap. 
In the former case, the dual constraints of $d_1$ and $d_2$ share the dual variable $\alpha(a_d)$.
In the latter case, both $d_1$ and $d_2$ belong to the same group $G^{(q)}_k$. 
Hence, the properties of layered decompositions imply that
one of the critical edges in $\pi(d_1)$ also appears in the $\mypath(d_2)$.
Thus, in either case, when $d_1$ is raised, the LHS of $d_2$ is also raised by an amount $\delta(d_1)$.
On the other hand, $d_2\in U_{i+1}$ and so, even after the above raise in the LHS value,
$d_2$ is still $(1-\xi^j)$-unsatisfied. As we are considering stage $j$,
all the demand instances in $G_k$ are $(1-(\xi)^{j-1})$-satisfied.
The gap between $(1-\xi^{j-1})p(d_2)$ and $(1-\xi^j)p(d_2)$ is $(\xi^{j-1} - \xi^j)p(d_2)$.
We see that even after the value of the LHS of the dual constraint of $d_2$ is raised by an amount $\delta(d_1)$,
the above gap is not bridged.  It follows that 
\[
(\xi^{j-1} - \xi^j) p(d_2) \geq \delta(d_1) \geq \frac{\xi^j\cdot p(d_1)}{(\Delta+1)}
\]
This implies that
\[
\frac{p(d_2)}{p(d_1)} \geq \frac{\xi}{(1-\xi)(\Delta + 1)}.
\]
We derive the claim by substituting $\xi=14/15$ and $\Delta=6$.

Consider any demand instance $d_L\in U_L$. There must exist a demand instance $d_{L-1}$ in $U_{L-1}$
such that $d_{L-1}$ kills $d_L$. In general, we can find a sequence of demand instances
$d_L, d_{L-1}, \ldots, d_1$ such that for $1\leq i\leq L-1$, $d_i$ kills $d_{i+1}$.
By the above claim, for $1\leq i\leq L-1$, $p(d_{i+1})\geq 2p(d_i)$.
It follows that $p(d_L)\geq 2^{L-1} p(d_1)$. 
Hence, $L \leq 1 + \log(p(d_L)/p(d_1)) = O(\log(p_{\max}/p_{\min}))$.
\qed

The properties of layered decomposition imply that the above two-phase algorithm satisfies the interference
property, governed by parameters $\Delta=6$ and $\lambda=(1-\epsilon)$.
Therefore, by Lemma \ref{lem:AAA}, it follows that the algorithm has an approximation ratio of $7/(1-\epsilon)$.
For $\epsilon'>0$, we can choose $\epsilon$ suitably and obtain an approximation ratio of $(7+\epsilon')$.
We have proved the main result of the paper.

\begin{theorem}
\label{thm:main}
There exists a distributed algorithm for the unit height case of the throughput maximization problem on tree-networks 
with approximation ratio $(7+\epsilon)$ and number of (communication) rounds is at most 
$O(\TMIS \log n \log(1/\epsilon) \log(p_{\max}/p_{\min}))$,
where $\epsilon>0$ is any constant.
\end{theorem}

\noindent
{\bf Remark: }Recall that Panconesi and Sozio \cite{Pancc} presented an algorithm for the unit height case of
line-networks. Their algorithm follows the two-phase framework with the slackness parameter $\lambda=1/(5+\epsilon)$.
On the other hand, our algorithm has $\lambda = (1-\epsilon)$.
A comparison of the two algorithms is in order. We reformulate their algorithm to suit our framework.
Their algorithm also classifies the demand instances into groups 
(based on length) and processes the groups in epochs. However, 
each epoch consists of only a single stage. They split the stage into multiple iterations/steps.
In any iteration, a demand instance $d$ which is $(1/(5+\epsilon))$-satisfied is ignored
for the rest of the first phase. In contrast, our algorithm works in multiple stages,
where in each stage, we make gradual progress towards making the demand instances
within the group to be $(1-\epsilon)$-satisfied. In particular,
in stage $j$, a demand instance which is $(1-\xi^j)$-satisfied is not ignored;
it exits the current stage, but it is included in the MIS computations in the next stage.
\\

\noindent
{\bf Distributed Implementation: }
Here, we sketch certain aspects of implementing the algorithm in a distributed manner.
Let $M_{\max}$ be the number of bits needed to encode the information about any demand 
(such as its end-points and profit).

For now, assume that the values $p_{\max}$ and $p_{\min}$ are known to all the processors.
Under this assumption, we can count the number of epochs, stages and iterations exactly.
The number of epochs is $2\ceil{\log n}$ (the maximum depth of ideal tree decompositions); 
the number of stages within each epoch is $\ceil{\log_{\xi} \epsilon}$, where $\xi=14/15$;
the number of iterations within each stage is $c\log (p_{\max}/p_{\min})$, where $c$ is a suitable constant.

Each processor $P$ computes the ideal tree-decomposition and the corresponding layered decomposition
for each tree-network $T$ accessible to it.
Each processor maintains the values of the dual variables correspoding to its demand instances.
The algorithm proceeds in a synchronous fashion consisting of multiple communication rounds,
where each round corresponds to a tuple $\langle f_1, f_2, f_3\rangle$,
where $f_1$, $f_2$ and $f_3$ are the epoch, stage and iteration number of the pseudocode.
Given a tuple $\langle f_1, f_2, f_3\rangle$,
a processor $P$ can determine which demand instances can participate in the MIS calculation of this 
communication round. The MIS calculation is performed considering the {\em conflict graph}:
the demand instances participating in the MIS computation form the vertices
and an edge is drawn between a pair of vertices, if they are conflicting.
The number of vertices is at most $N=mr$, where $m$ is the number of demands
and $r$ is the number of tree-networks.
The MIS can computed using either the randomized algorithm of Luby \cite{Luby}
or using the deterministic procedure of network decompositions \cite{PancSrini}.
In the former case, the number of (communication) rounds needed in $O(\log N)$, whereas
in the latter case, it is $O(2^{\sqrt{\log N}})$.
Each processor $P$ contains at most one demand instance $d$ belonging to the MIS.
The processor $P$ raises the dual variables corresponding to $d$ as given in the pseudocode.
The new dual variables are transmitted to the processors sharing a common resource with $P$.
Upon receiving the new dual variables, each processor updates the dual variables of its demand instances suitably.
Each processor $P$ raises at most a constant number of dual variables in each iteration (because the critical set 
size $\Delta=6$ is a constant) and the amount of increase is at most $p_{\max}$.
Thefore, the message size is bounded by $M_{\max}$.
The stack is implemented in a distributed manner. Each processor $P$ maintains its own stack.
Whenever the processor $P$ raises a demand instance $d$, it pushes $d$ onto its stack
along with a tuple $t$ (consisting of the corresponding epoch, stage and iteration numbers).

The second phase proceeds in the reverse order. Each round of the second phase
is associated with a tuple $t$ (consisting of epoch, stage and iteration numbers).  
A processor $P$ will compare tuple $t'$ on the top of the stack
with the tuple $t$; if they match, then it will pop the demand instance $d$ on the top of the stack.
It will output $d$, if feasibility is maintained.
In this case, the processor $P$ will inform its neighboring processors
that $d$ has been included in the output solution.

Finally, we note that it is not difficult to bypass the assumption that all the processors know 
the values of $p_{\max}$ and $p_{\min}$.

\section{Arbitrary Height Case for Tree-Networks}
\label{sec:arbit}
Panconesi and Sozio \cite{Pancc} presented a distributed $(20+\epsilon)$-approximation algorithm
for the unit height case of the line-networks. Then, they extended this algorithm 
for the arbitrary height case and obtained a $(55+\epsilon)$-approximation algorithm \cite{Pancj}.
In this section, we extend our $(7+\epsilon)$-approximation algorithm for the unit height case
of the tree-networks to the arbitrary height case and derive a $(20+\epsilon)$-approximation algorithm. 
We note that the extra ideas needed for the extension roughly follow the theme of Panconesi and Sozio \cite{Pancj}. 
The main difference is that their algorithm follows the two-phase framework with the 
slackness parameter being $\lambda = 1/(5+\epsilon)$, whereas we aim for $\lambda=(1-\epsilon)$.
Below we highlight the necessary changes.

We first develop some notation for the arbitrary height case.
The problem setup is as before, except that each demand $a\in \calA$
also has a bandwidth requirement (or {\em height}) $h(a)$.
All the edges in all the tree-networks are assumed to provide a uniform bandwidth of $1$ unit.
For each demand $a\in \calA$ owned by a processor $P$ and each tree-network $T$ accessible to $P$,
create a copy of $a$ with the same end-points, height and profit; these copies are called
demand instances of $a$.
Let $\calD$ denote the set of all demand instances (over all demands).
A demand instance $d\in \calD$ belonging to a tree-network $T$ has a height $h(d)$ and a profit $p(d)$,
and it can be viewed as a path between its end-points in $T$.
A feasible solution selects a subset of demand instances $S\subseteq \calD$
such that: (i) for any demand $a\in \calA$, at most one demand instance of $a$ is selected;
(ii) for any edge $e$ in some tree-network $T$,
among the demand instances in $S$, the sum of heights of the demand instances passing through $e$ is at most $1$ unit.
The profit of the solution is the sum of profits of the demand instances contained in it.
The goal is to choose the solution having the maximum profit.

We classify the input demand instances $d\in \calD$ into two categories based on their height:
(i) {\em narrow instances: } $d$ is said to be narrow, if $h(d)\leq 1/2$;
(ii) {\em wide instances: } $d$ is said to be wide, if $h(d)>1/2$.
Notice that two wide demand instances which are overlapping cannot be picked by a feasible solution.
Hence, if our input consists only of wide demand instances, we can reuse the algorithm for the unit height case
and get a $(7+\epsilon)$ approximation factor.
We next describe a $(13+\epsilon)$-approximation algorithm for the special case where in the input
consists of only narrow instances. The final algorithm will be derived by combining the above two algorithms.

\subsection{Narrow Instances}
Here, we assume that the input consists of only narrow instances and develop a $(13+\epsilon)$-approximation algorithm.

\subsubsection*{LP and Dual}
The LP and the dual have to be modified suitably to reflect the notion of heights.
Recall that for each demand instance $d\in \calD$, we have a variable $x(d)$.
\begin{eqnarray*}
\max \quad & & \sum_{d \in {\calD}} x(d) \cdot p(d) \\
\sum_{d\in \calD ~:~ d \sim  e} x(d) \cdot h(d)& \leq & 1 \quad \quad \mbox{for edges $e\in \calE$} \\
\sum_{d \in \Inst(a)} x(d) & \leq & 1 \quad \quad \mbox{for all demands $a\in \calA$} \\
x(d) & \geq & 0 \quad \quad \mbox{for all demand instances $d\in \calD$}
\end{eqnarray*}
The first set of constraints capture the fact that the cumulative height of the demand instances
active on an edge $e$ cannot exceed one unit.
Similarly, the second set of constraints capture the fact that a feasible solution
can select at most one demand instance belonging to any demand.

The dual of the LP is as follows.
For each demand $a\in \calA$ and each edge $e\in \calE$, the dual includes a variable $\alpha(a)$ and $\beta(e)$,
respectively. Similarly, for each demand instance $d\in \calD$, 
the dual includes a constraint; we call this the {\em dual constraint of $d$}.
Recall that $a_d$ denotes the demand to which a demand instance $d$ belongs.
\begin{eqnarray}
\nonumber
\min \quad & & \sum_{a \in {\calA}} \alpha(a) + \sum_{e\in \calE} \beta(e) \\
\nonumber
\alpha(a_d) + h(d) \sum_{e~:~ d \sim  e} \beta(e) & \geq & p(d) \quad \quad \mbox{for demand instances $d\in \calD$} 
\end{eqnarray}
The dual also consists of the non-negativity constraints: for all $a\in \calA$ and $e\in \calE$, $\alpha(a)\geq 0$
and $\beta(e)\geq 0$. 

\subsubsection*{Two-phase Framework}
We modify the two-phase framework as follows.
As before, the algorithm proceeds iteratively and a typical iteration is performed as below. 
We choose suitable independent set $I$ and raise each demand instance $d\in I$.
The slackness computation is modified to reflect the notion of heights in the constraints.
Define the slackness to be:
\[
s = p(d) - \left( \alpha(a_d) + h(d)\cdot \sum_{e~:~d\sim e} \beta(e) \right).
\]
We next select a suitable subset $\pi(d)$ consisting of critical edges on which $d$ is active.
The strategy for raising the dual variables is modified slightly.
We raise $\alpha(a_d)$ by $\delta(d)$ and for each $e\in \pi(d)$, 
raise $\beta(e)$ by $2|\pi(d)|\delta(d)$. Towards that goal, define $\delta(d)=s/(1+2h(d)|\pi(d)|^2)$.
We see that the dual constraint is satisfied tightly in the process.
The second phase of the algorithm remains the same.

The parameters critical set size $\Delta$ and slackness $\lambda$ are defined as before.
Similarly, the concept of interference property remains the same.
Lemma \ref{lem:AAA} can be extended as follows, using similar arguments.

\subsubsection*{Approximation Guarantee}
\begin{lemma}
\label{lem:JJJ}
Suppose the input consists of only narrow instances.
Consider any algorithm satisfying the interference property and governed by parameters $\Delta$ and $\lambda$.
Then the feasible solution $S$ produced by the algorithm satisfies 
$p(S)\geq \left(\frac{\lambda}{1+2\Delta^2}\right)\cdot p(\Opt)$.
\end{lemma}
\proof
As before, let $\alpha(\cdot)$ and $\beta(\cdot)$ be the dual variable assignment produced at the end of the first phase.
Convert the assignment $\langle \alpha, \beta\rangle$ into a dual feasible solution
by scaling the values by an amount $1/\lambda$:
for each demand instance $d$, set $\wh{\alpha}(d) = \alpha(d)/\lambda$ and 
for each edge $e$, set $\wh{\beta}(e)=\beta(e)/\lambda$.
Notice that the $\langle \wh{\alpha},\wh{\beta}\rangle$ forms a feasible dual solution.

Let $\val(\alpha, \beta)$ and $\val(\wh{\alpha},\wh{\beta})$ be the objective
value of the dual assignment $\langle \alpha, \beta \rangle$
and the dual feasible solution $\langle \wh{\alpha},\wh{\beta}\rangle$,
respectively. By the weak duality theorem, $\val(\wh{\alpha},\wh{\beta}) \geq p(\Opt)$.
The scaling process implies that $\val(\wh{\alpha},\wh{\beta}) = \val(\alpha,\beta)/\lambda$.
We now establish a relationship between $\val(\alpha,\beta)$ and $p(S)$.

Let $R$ denote the set of all demand instances that are raised in the first phase.
The value $\val(\alpha,\beta)$ can be computed as follows.
For any $d\in R$, the variable $\alpha(a_d)$ is raised by an amount $\delta(d)$
and for each $e\in \pi(d)$, the variable $\beta(e)$ is raised by an amount $2|\pi(d)|\delta(d)$.
We have that $|\pi(d)|\leq \Delta$.
So, whenever a demand instance $d\in R$ is raised, the objective value raises at most $(2\Delta^2+1)\delta(d)$.
Therefore,
\begin{eqnarray}
\label{eqn:HHH}
\val(\alpha,\beta) &\leq& (2\Delta^2+1)\sum_{d\in R} \delta(d).
\end{eqnarray}

We next compute the profit of the solution $p(S)$.
For a pair of demand instances $d_1, d_2\in R$, we say that $d_1$ is a {\em predecessor} of $d_2$,
if the pair is conflicting and $d_1$ is raised before $d_2$;
in this case $d_2$ is called the {\em successor} of $d_1$.
For a demand instance $d\in R$, let $\mypred(d)$ and $\mysucc(d)$ denote the set of predecessors and successors of $d$,
respectively; we exclude $d$ from both the sets.

Define $K=R-S$; we say that demand instances in $K$ are {\em killed} by the procedure.
We say that a demand $d'$ is killed by a demand $d$ if the following three conditions are true: (i) $d\in S$;
(ii) both $d$ and $d'$ belong to the same demand;  (iii) $d'\in \mypred(d)$.
For a demand $d\in S$, let $K_1(d)$ denote the set of demand instances killed by $d$; 
notice that $K_1(d)\subseteq \mypred(d)$.
Let $K_1$ be the union of $K_1(d)$ over all demand instances $d\in S$.
Let $K_2$ be the demand instances in $K$ that could not be added to the solution because of bandwidth constraints
(i.e., $K_2$ consists of demand instances $d$ such that for some edge $e\in \mypath(d)$, 
$h(d)+\sum_{d'\in S~:~d'\sim e} h(d') > 1$).

Consider any demand instance $d\in S$ and we shall computer a lowerbound on $p(d)$.
Consider the iteration in which $d$ was raised; after the raise, the LHS of the constraint
of $d$ becomes equal to the RHS (i.e., $p(d)$). 
The demand instance $d$ contributes at least $\delta(d)$ to the LHS.
In the previous iterations, any demand instance $d'\in K_1(d)$ would have contributed $\delta(d')$ to the LHS.
Similarly, for any $e\in \mypath(d)$ and any demand instance $d'\in \mypred(d)\cap K_2$ such that $e\in \pi(d')$,
$d'$ would have contributed $2h(d)|\pi(d')|\delta(d')$ to the LHS. Thus, we see that:
\begin{eqnarray*}
p(d) 
\geq 
\delta(d) 
+ \left(\sum_{d'\in K_1(d)} \delta(d')\right) 
+ \left(\sum_{e\in \mypath(d)} \quad \sum_{d'\in \mypred(d)\cap K_2~:~e\in \pi(d')} 2h(d)|\pi(d')|\delta(d')\right).
\end{eqnarray*}
Summing up $p(d)$ over all the demand instances $d\in S$, we get that
\begin{eqnarray*}
p(S) 
\geq 
\sum_{d\in S} \delta(d) 
+ \left(\sum_{d\in S} \quad \sum_{d'\in K_1(d)} \delta(d')\right) 
+ \left(\sum_{d\in S} \quad \sum_{e\in \mypath(d)} \quad \sum_{d'\in \mypred(d)\cap K_2~:~e\in \pi(d')} 2h(d)|\pi(d')|\delta(d')\right).
\end{eqnarray*}
Rewriting the second and the third terms of the RHS:
\begin{eqnarray}
\label{eqn:dxy}
\sum_{d\in S} p(d) 
\geq 
\sum_{d'\in S} \delta(d') 
+ \left(\sum_{d'\in K_1} \delta(d')\right)
+ \left(\sum_{d'\in K_2} 2 |\pi(d')|\delta(d') \sum_{e\in \pi(d')} \quad \sum_{d\in S\cap \mysucc(d')~:~d\sim e} h(d)\right).
\end{eqnarray}
Now let us analyze the third term in the RHS.
Consider any $d'\in K_2$.
The demand instance $d'$ could not be added to $S$ because of the bandwidth constraint being
violated at some edge $e'\in \mypath(d')$. Meaning,
\[
h(d') + \sum_{d\in S\cap \mysucc(d')~:~d\sim e'} h(d) > 1.
\]
Since all demand instances are assumed to be narrow, we have that $h(d')\leq 1/2$. 
It follows that 
\[
\sum_{d\in S\cap \mysucc(d')~:~d\sim e'} h(d) > 1/2.
\]
Let $X=\{d\in S\cap \mysucc(d')~:~d\sim e'\}$.
By the interference property, all the demand instances in $X$ are active
at one of the edges in $\pi(d')$.
It follows that there exists an edge $\wt{e}\in \pi(d')$ such that 
\[
\sum_{d\in \wt{X}} h(d) > \frac{1}{2|\pi(d')|},
\]
where $\wt{X}=\{d\in X~:~d\sim \wt{e}\}$.
Consider the third term in the RHS of the formula (\ref{eqn:dxy}).
The second summation is over all the edges $e\in \pi(d')$.
We replace this summation by the single quantity corresponding to $\wt{e}$.
We get that
\begin{eqnarray*}
\sum_{d\in S} p(d) 
\geq
\sum_{d'\in S} \delta(d') 
+ \left(\sum_{d'\in K_1} \delta(d')\right)
+ \left(\sum_{d'\in K_2} 2 |\pi(d')|\delta(d') \quad \sum_{d\in S\cap \mysucc(d')~:~d\sim \wt{e}} h(d)\right)
\end{eqnarray*}
Therefore,
\begin{eqnarray*}
\sum_{d\in S} p(d) 
&\geq& \left(\sum_{d'\in S} \delta(d')\right) + \left(\sum_{d'\in K_1} \delta(d')\right) 
+ \left(\sum_{d'\in K_2} 2 |\pi(d')|\delta(d') \times \frac{1}{2|\pi(d')|}\right)\\
&\geq& \sum_{d' \in R} \delta(d').
\end{eqnarray*}
The second statement follows from the fact that $R$ is the union of $S$, $K_1$ and $K_2$.
Comparing (\ref{eqn:HHH}), we see that $\val(\alpha,\beta)\leq (2\Delta^2+1)p(S)$.
The lemma is proved by the observations made at the beginning of the proof.
\qed

\subsubsection*{Distributed Algorithm}
Fix any $\epsilon > 0$. 
Our goal is to design a two-phase procedure with critical set size $\Delta=6$ and slackness parameter 
$\lambda=(1-\epsilon)$.
The distributed algorithm is similar to that of the unit height case (Section \ref{sec:distributed}).
We use the same layered decompositions given by Lemma \ref{lem:DDD} (with $\Delta=6$ and length $\ell=\log n$).
The only difference is that we set the parameter $\xi=\frac{c}{c+\hmin}$, for some suitable constant $c$.
Arguments similar that of Lemma \ref{lem:EEE} can be used to show that in any epoch $k$ and any stage $j$,
the algorithm takes at most $O(\log (p_{\max}/p_{\min}))$ steps (or iterations).
The number of epochs in $O(\log n)$, as before.
The number of stages within each epoch is $\log_{\xi}(\epsilon)$.
The above value of $\xi$ ensures that the above quantity is at most $O(1/\hmin)\log(1/\epsilon)$.
Therefore, the total number of steps is 
$O(\TMIS \cdot (1/h_{\min}) \cdot (\log n) \cdot \log(1/\epsilon) \cdot \log(p_{\max}/p_{\min}))$.
The above algorithm satisfies the interference property with critical set size $\Delta=6$ and slackness
parameter $\lambda = (1-\epsilon)$. By Lemma \ref{lem:JJJ}, 
the approximation ratio is $(2\Delta^2+1)/\lambda = 73/(1-\epsilon)$.
We have established the following lemma.

\begin{lemma}
\label{lem:KKK}
Fix any $h_{\min}\leq 1/2$ and $\epsilon>0$. Consider the special case of the scheduling problem on tree-networks
with heights wherein all the demands $a$ are narrow and satisfy $h(a)\geq h_{\min}$.
There exists a distributed algorithm for the above problem with approximation ratio $(73+\epsilon)$.
The number of (communication) rounds is at most 
$O\left( \frac{\TMIS}{h_{\min}} \log n \log{ \frac{1}{\epsilon} } \log{ \frac{p_{\max}}{p_{\min}} }\right)$.
\end{lemma}

\subsection*{Overall Algorithm}
Fix any $\epsilon'>0$. We present an algorithm within an approximation ratio of $(80+\epsilon')$,
for the arbitrary height case of tree-networks.

We classify the demand instances into wide and narrow instances.
Let $\Opt_1$ and $\Opt_2$ denote the optimal solutions considering only the wide and narrow instances,
respectively. Notice that $p(\Opt) \leq p(\Opt_1) + p(\Opt_2)$.
For the wide instances, we run the algorithm for the unit height case (Theorem \ref{thm:main})
and obtain a solution $S_1$ such that $p(\Opt_1)\leq (7+\epsilon')p(S_1)$.
Next we run the algorithm for the narrow instances (Lemma \ref{lem:KKK})
and obtain a solution $S_2$ such that $p(\Opt_2)\leq (73+\epsilon')p(S_2)$.
Output a combined solution $S$ as follows.
For each tree-network $T\in \calT$, consider the set of demand instances scheduled on $T$ by the solution $S_1$
and the set of demand instances scheduled on $T$ by the solution $S_2$; among the two sets, choose the one
with the higher profit. It is easy to see that the bandwidth constraints are satisfied by $S$.
Furthermore, for any demand $a$, either all the demand instances of $a$ are narrow
or all of them are wide. Therefore, $S$ will pick at most one demand instance from any demand.
This shows that $S$ is indeed a feasible solution.
We have that $p(S)\geq \max\{p(S_1),p(S_2)\}$. It follows that $p(\Opt)\leq (80+2\epsilon')p(S)$.
We have established the following theorem.

\begin{theorem}
\label{thm:LLL}
Fix any $h_{\min}\leq 1/2$ and $\epsilon>0$. Consider the scheduling problem on tree-networks
with arbitrary heights wherein all the demands $a$ satisfy $h(a) \geq h_{\min}$.
There exists a distributed algorithm for the above problem with approximation ratio $(80+\epsilon)$.
The number of (communication) rounds is at most 
$O\left( \frac{\TMIS}{h_{\min}} \log n \log{ \frac{1}{\epsilon} } \log{ \frac{p_{\max}}{p_{\min}} }\right)$.
\end{theorem}

\section{Line-Networks}
Recall that Panconesi and Sozio \cite{Pancj} presented distributed algorithms
for the case of line-networks with windows. For the unit height case the approximation ratio was $(20+\epsilon)$
and for the arbitrary height case, the ratio was $(55+\epsilon)$. 
In this section, we obtain a improved approximation ratios $(4+\epsilon)$ and $(23+\epsilon)$, respectively.
We next explain the new algorithms within our framework.

We first develop some notation. Let $\calP$ be the set of $m$ processors, $\calT$ be the set of $r$ resources
and $\calA$ be the set of $n$ demands.
We divide timeline into $n$ discrete timeslots, $1,2,\ldots, n$. Each processor $P$ owns a demand $a\in \calA$.
Each demand $a\in \calA$ is specified by a window $[\rt(a),\dl(a)$ and a processing time $\rho(a)$,
where $\rt(a)$ and $\dl(a)$ are the release time and deadline of $a$.
A profit $p(a)$ and a height $h(a)$ are associated with each demand $a$. 
Consider a processor $P$ and the demand $a$ owned by $P$.
For each resource $T$ accessible by $P$ and each interval of length $\rho(a)$ contained within $[\rt(a), \dl(a)]$,
create a demand instance $d$; its profit and height are the same as that of $a$;
the number of demand instances is at most $n|\Acc(P)|$, where $\Acc(P)$
is the set of resource accessible to $P$. Let $\calD$ denote the set of all demand instances.
Each demand instance $d\in \calD$ is described by a starting time $s(d)$, and ending time $e(d)$,
a profit $p(d)$, a height $h(d)$ and the resource to which it belongs.
Recall that the time-line can be viewed as a tree-network with $n+1$ vertices.
In other words, the case of line-networks can be reduced to the case of tree-networks.
Therefore, Theorem \ref{thm:main} and \ref{thm:LLL} apply to the case of line-networks.
We next show how to improve these results in the case of line networks.
The improvements are obtained by designing layered decompositions with better parameters.

\subsubsection*{Improved Layered Decomposition}
In the case of tree-networks, we derived decompositions with critical set size $\Delta=6$
and length $\ell= O(\log n)$. In the case special of line-networks, we show how to construct
decompositions with parameters $\Delta=3$ and $\ell = O(\log(L_{\max}/L_{\min}))$ ($L_{\max}$ and $L_{\min}$
are the maximum and minimum length of the demand instances).
We note that this decomposition is implicit in \cite{Pancj}.

Partition the demand instances in to $\ell=\ceil{\log(L_{\max}/L_{\min})}$ categories based
on their length, where the length of a demand instance is $\len(d)=e(d)-s(d)+1$.
For $1\leq i\leq \ell$, define $G_i = \{d~:~2^{i-1} L_{\min} \leq \len(d) \leq 2^i L_{\min}$.
Define an ordering $\sigma=G_1, G_2, \ldots, G_{\ell}$.
For each demand instance $d$, let $\mymid(d)=\floor{(s(d)+e(d))/2}$ be the mid-point of $d$.
Define $\pi(d)=\{s(d),\mymid(d),e(d)\}$, for all $d\in \calD$.
It is not difficult to argue that the pair $\pair{\sigma}{\pi}$ forms layered decomposition.

\subsubsection*{Unit Height Case}
Fix $\epsilon>0$. We modify the distributed algorithm given in Section \ref{sec:distributed}
to use the above layered decomposition. We suitably change the value of $\xi$ to 
$8/9$ (instead of $14/15$). This algorithm would satisfy the interference property with $\Delta=3$
and $\lambda=(1-\epsilon)$. So, the approximation ratio is $4/(1-\epsilon)$.
The number of (communication) 
rounds is $O(\TMIS \cdot \log(1/\epsilon) \cdot \log(L_{\max}/L_{\min})\cdot \log(p_{\max}/p_{\min})$.
We have established the following result.

\begin{theorem}
\label{thm:MMM}
There exists a distributed algorithm for the scheduling problem for the unit height case of line-networks with windows
with approximation ratio $(4+\epsilon)$. The number of (communication) rounds is at most 
$O\left( \TMIS \log{ \frac{1}{\epsilon} } \log \frac{L_{\max}}{L_{\min}} \log{ \frac{p_{\max}}{p_{\min}} }\right)$,
\end{theorem}

\subsubsection*{Arbitrary Height Case}
Here, we discuss the arbitrary height case. As in Section \ref{sec:arbit}, we partition
the set of demand instances into narrow and wide categories.
For the case of wide instances, Theorem \ref{thm:MMM} applies and 
yields an algorithm with an approximation ratio of $(4+\epsilon)$.
For the case of narrow instances, the algorithm is similar to that of Lemma \ref{lem:KKK};
the only change is that we set the parameter $\xi=\frac{c'}{c'+\hmin}$, for a suitable constant $c'$.
This way we get an algorithm with an approximation ratio of $19+\epsilon$ 
(because, in the current setup $\Delta=3$).
We obtain an overall algorithm by combining the solutions output by the above two algorithm;
the idea is same as that of Theorem \ref{thm:LLL}. We have established the following result.

\begin{theorem}
Fix any $h_{\min}\leq 1/2$ and $\epsilon>0$. Consider the scheduling problem on tree-networks
with arbitrary heights wherein all the demands $a$ satisfy $h(a) \geq h_{\min}$.
There exists a distributed algorithm for the above problem with approximation ratio $(23+\epsilon)$.
The number of (communication) rounds is at most 
$O\left( \frac{\TMIS}{h_{\min}} \log n \log{ \frac{1}{\epsilon} } \log{ \frac{p_{\max}}{p_{\min}} }\right)$.
\end{theorem}

\bibliographystyle{plain}
\bibliography{main}

\appendix
\section{A Sequential Algorithm for Tree-networks}
\label{sec:Lewin}
Here, we present a sequential algorithm satisfying the interference property with 
critical set size $\Delta=2$ and slackness parameter $\lambda = 1$.

Let $\calT=\{T_1, T_2, \ldots, T_r\}$ be the given set of tree-networks.
For $1\leq i\leq r$, construct a rooted tree $H_i$ by arbitrarily selecting a node $g_i\in V$ 
and making $g_i$ as the root of $T_i$. Let $\calH = \{H_1, H_2, \ldots, H_r\}$
be the set of rooted trees constructed by the above process.

Consider an input tree-network $T\in \calT$. 
Let $H\in \calH$ be the rooted tree corresponding to $T$ with $g$ as the root.
For a node $x$, define its {\em depth} (or height) to be the number
of the nodes along the path from $g$ to $x$; the root $g$ itself is defined to have to depth $1$.
For a demand instance $d\in \calD(T)$, let $\mu(d)$ denote the node in $H$
having the least depth among all the nodes appearing in $\mypath(d)$.
We say that $d$ is {\em captured} at the node $\mu(d)$; observe that $\mu(d)$ is uniquely determined.
See Figure \ref{fig:5}. A rooted-tree $H$ has been constructed by picking the node $1$ as the root.
The demand instance $d=\pair{4}{13}$ will be captured at the node $\mu(d)=2$.
With respect to $H$, a node $y$ is said to be an {\em ancestor} of $x$,
if $y$ appears along the path from $g$ to $x$; in this case, $x$ is said to be a {\em descendent} of $y$.
By convention, we do not consider $x$ to be an ancestor or descendent of itself.

Consider a demand instance $d\in \calD(T)$. 
We have that $\mypath(d)$ passes through the node $\mu(d)$. 
Let $\pi(d)$ denote the set of edges of $\mypath(d)$ that are adjacent to $\mu(d)$.
If $\mu(d)$ is one of the end-points of $d$ then $\pi(d)$ will have only one edge;
otherwise, it will have two edges. 
See Figure \ref{fig:5}. The set $\pi(d)$ for the demand instance $d=\pair{4}{13}$
is given by $\pi(d)=\{\pair{2}{4},\pair{2}{5}\}$.
We can now make the following important observation.

\begin{figure}[t!]
\begin{center}
\begin{boxedminipage}{\hsize}
\begin{small}
\begin{tabbing}
xx\=xx\=xx\=xx\=xxx\=xxx\=\kill
\textbf{Begin}  \\
\> // Initialize \\
\> For all $a\in \calA$, set $\alpha(a)=0$; for all $e\in \calE$, set $\beta(e)=0$.\\
\> Initialize an empty stack.\\
\>\\
\> // First phase\\
\> For $i$ = $1$ to $r$\\
\> \> While\\
\> \> \>Let $U=\{d\in \calD(T_i)~:~\mbox{dual constraint of $d$ is unsatisfied}\}$.\\
\> \> \>If $U=\emptyset$, exit the loop.\\
\> \> \>Let $d$ be the element of $U$ appearing earliest in $\sigma(T_i)$.\\
\> \> \>Compute slackness: $s=p(d) - \alpha(a_d) - \sum_{e:d\sim e} \beta(e)$.\\
\> \> \>Compute: $\delta(d) = s/(|\pi(d)|+1)$.\\
\> \> \>Raise the variables: $\alpha(a_d) \leftarrow \alpha(a_d)+\delta(d)$; 
          for all $e\in \pi(d)$, $\beta(e) \leftarrow \beta(e)+\delta(d)$.\\
\> \> \>Push $d$ into the stack.\\
\>\\
\> // Second Phase\\
\> \> $S=\emptyset$.\\
\> \> While(stack not empty)\\
\> \> \> Pop the top element $d$ of the stack\\
\> \> \> If $S\cup \{d\}$ is an independent set, then add $d$ to $S$.\\
\> Output $S$.\\
\textbf{End}
\end{tabbing}
\end{small}
\end{boxedminipage}
\end{center}
\caption{Sequential Algorithm: Pseudocode}
\label{fig:Lewin-pseudo}
\end{figure}

\begin{observation}
\label{obs:AAA}
Consider any two overlapping demand instances $d_1, d_2\in \calD(T)$.
If $d_1$ is captured at a node $z$ and $d_2$ is captured at an ancestor of $z$,
then $\mypath(d_2)$ will include one of the edges from $\pi(d_1)$.
Furthermore, if $d_1$ and $d_2$ are captured at the same node $z$,
then also $\mypath(d_2)$ will include one of the edges from $\pi(d_1)$.
\end{observation}

Based on the above observation, our algorithm works as follows.
In the two-phase framework, in each iteration, we need to determine an independent set $I$.
We shall perform the above task by selecting a singleton demand instance $d$ (forming a trivial independent set);
we can afford to do this, since we are designing a sequential algorithm.
Towards determining $d$, we define an ordering of the demand instances for each tree-network.
Consider a tree-network $T\in \calT$ and let $H$ be the rooted tree corresponding to $T$ with $g$ as the root. 
Order the demand instances $d\in \calD(T)$ in the descending order of the depth of $\mu(d)$;
ties are broken arbitrarily. Thus, demand instances captured at the bottom-most leaves of $H$ will 
be placed first and those captured at the root will be placed last.
Let $\sigma(T)$ denote the ordering obtained by the above process.

The first phase works in $r$ rounds, where the $i$th round will process the tree-network $T_i$.
The round $i$ works in multiple iterations, where each iteration is performed as described below.
Let $U$ be the demand instances belonging to $T_i$ whose constraint is unsatisfied (recall that $\lambda=1$).
Among demand instances in $U$, pick the demand instance $d$ appearing earliest in the ordering $\sigma(T_i)$
and raise the demand $d$: namely, raise the dual variable $\alpha(a_d)$ and the dual variables
of the edges found in $\pi(d)$. The $i$th round is completed when the dual constraints of 
all the demand instances of $\calD(T_i)$ are satisfied.
The pseudocode of the algorithm is presented in Figure \ref{fig:Lewin-pseudo}.

Observation \ref{obs:AAA} shows that the algorithm satisfies the interference property.
Furthermore, for any demand instance $d$, $|\pi(d)|\leq 2$.
Thus, we have a two-phase procedure with parameters $\Delta=2$ and $\lambda=1$.
Lemma \ref{lem:AAA} shows that the algorithm is a $3$-approximation algorithm.
In the special case where there is only one tree-network, we do not need to raise
the dual variables $\alpha(\cdot)$ and the approximation ratio can be improved to $2$.
The resulting algorithm would essentially be the same as that of Lewin-Eytan et al. \cite{Lewin-Eytan}.
The round complexity of the algorithm can be as high as $n$, since only a single demand instance is 
raised in each iteration.

\end{document}